\documentclass[journal=jctcce,manuscript=article]{achemso}
\usepackage[utf8]{inputenc}
\usepackage{chemformula} % Formula subscripts using \ch{}
\usepackage[T1]{fontenc} % Use modern font encodings
\usepackage{hyperref}
\usepackage{xcolor}
\usepackage{tabularx}
\usepackage{soul}
\usepackage{threeparttablex} % table with notes
\usepackage{longtable} % table
\usepackage{rotating} % rotating tables and figures
\usepackage{amsmath}
\usepackage{comment}
\usepackage{gensymb}
\usepackage{float}
\usepackage[version=4]{mhchem}
\usepackage[normalem]{ulem}

\setkeys{acs}{maxauthors=10}
\setkeys{acs}{etalmode=truncate}
\SectionNumbersOn % for section numbers
\usepackage{xr}
\makeatletter
\newcommand*{\addFileDependency}[1]{% argument=file name and extension
\typeout{(#1)}
\@addtofilelist{#1}

\IfFileExists{#1}{}{\typeout{No file #1.}}
}\makeatother

\newcommand*{\myexternaldocument}[1]{%
\externaldocument{#1}%
\addFileDependency{#1.tex}%
\addFileDependency{#1.aux}%
}
\myexternaldocument{si-revised}

\author{Praveen Muralikrishnan}
\affiliation[CEMS]{Department of Chemical Engineering and Materials Science, University of Minnesota, Minneapolis, MN 55455, USA}
\alsoaffiliation[CTC]{Chemical Theory Center, University of Minnesota, Minneapolis, MN 55455, USA}

\author{Jonathan W. P. Zajac}
\affiliation[UMN]{Department of Chemistry, University of Minnesota, Minneapolis, MN 55455, USA}
\alsoaffiliation[CTC]{Chemical Theory Center, University of Minnesota, Minneapolis, MN 55455, USA}

\author{Caryn L. Heldt}
\affiliation[MTU]
{Department of Chemical Engineering, Michigan Technological University, Houghton, MI 49931, USA}

\author{Sarah L. Perry}
\affiliation[UMA]
{Department of Chemical Engineering, University of Massachusetts Amherst, MA 01003, USA}

\author{Sapna Sarupria}
\email{sarupria@umn.edu}
\affiliation[UMN]{Department of Chemistry, University of Minnesota, Minneapolis, MN 55455, USA}
\alsoaffiliation[CTC]{Chemical Theory Center, University of Minnesota, Minneapolis, MN 55455, USA}

\title[An \textsf{achemso} demo]
  {Thermodynamic Basis of Sugar-Dependent Polymer Stabilization: Informing Biologic Formulation Design}

\abbreviations{IR,NMR,UV}

\begin{document}

\begin{tocentry}

\begin{figure}[H]
 \centering
 \includegraphics[width=1\textwidth]{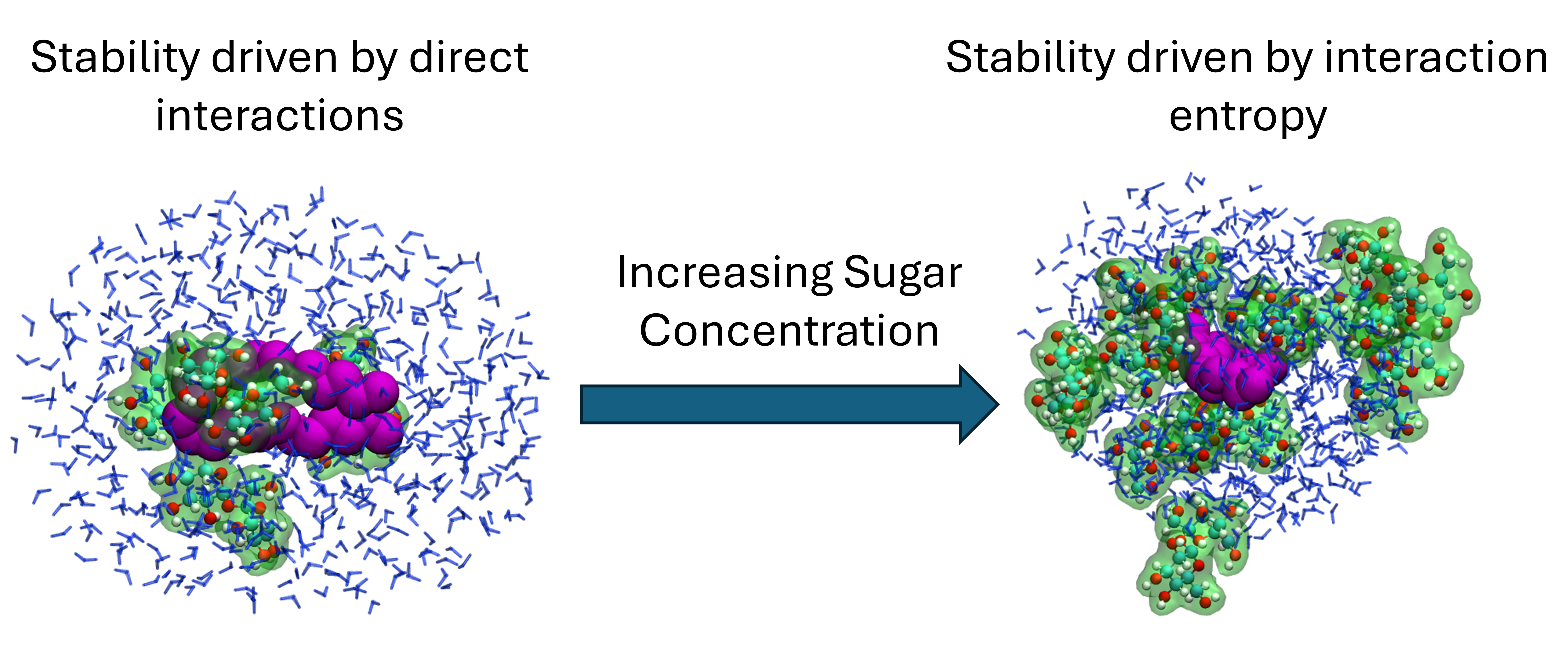}
    \label{fig:toc}
\end{figure}

\end{tocentry}

\begin{abstract}

The stabilization of macromolecules is fundamental to developing biological formulations, such as vaccines and protein therapeutics. In this study, we employ coarse-grained polymer models to investigate the impact of four sugars: $\alpha$-glucose, $\beta$-fructose, trehalose, and sucrose on macromolecule stability. Free energy decomposition and preferential interaction analysis indicate that polymer-sugar interactions favor folding at low concentrations while driving unfolding at higher concentrations. In contrast, the polymer-solvent soft interaction entropy consistently favors unfolding across all sugar concentrations under study. At low sugar concentrations, polymer-solvent interactions predominantly govern stabilization, whereas at higher concentrations, entropic penalties dictate polymer stability. Local mixing entropy demonstrates that binary sugar mixtures introduce entropic contributions that preferentially stabilize the folded state. These findings contribute to a more nuanced understanding of sugar-based excipient stabilization mechanisms, offering guidance for the rational design of stable biological formulations.
\end{abstract}

\section{Introduction}

Therapeutic protein and viral formulations (such as vaccines and gene therapy vectors) face significant stability challenges, requiring stringent storage conditions for long-term preservation.\cite{kroger2013general,centers2003guidelines, kumru2014vaccine, chen2009opportunities} These stability constraints necessitate a complex supply chain, known as the cold chain, to ensure the safe transport of formulations from manufacturers to patients. Disruptions in the cold chain can lead to protein degradation, compromising efficacy and limiting global accessibility.\cite{dai2021complexity, yu2021grand, fahrni2022management, pambudi2022vaccine} This underscores the urgent need for robust formulations that maintain stability under stress conditions, such as temperature fluctuations, pH changes, and mechanical agitation during manufacturing and storage.
 
A number of strategies have been explored to improve the stability of protein formulations, including site-specific mutagenesis, \cite{kamal2011vitro, fu2010increasing, hu2010role, tadokoro2013investigating} protein fusion, \cite{kolate2014peg, perezgasga2012substitution, rodriguez2008stabilization, da2010biochemical, budhavaram2010protein, cordes2012selective} and the incorporation of excipients (i.e., inactive stabilizing agents) in formulations. \cite{wang2015advanced, castaneda2022alternative, d2021development, peletta2023meeting} Among these strategies, excipients are the most widely adopted because they do not require protein-specific modification. \cite{kumru2014vaccine, zarzar2023high, wang2015advanced} Excipients used in protein-based therapeutics belong to diverse molecular families, including polysorbates, surfactants, carbohydrates, amino acids, synthetic amphiphilic polymers, and ionic liquids. \cite{castaneda2022alternative, kamerzell2011protein, chaudhari2012pharmaceutical, pockle2023comprehensive, patel2020pharmaceutical, shah2021pharmaceutical} These excipients primarily function to prevent physical degradation processes such as denaturation, aggregation, and surface adsorption, as well as chemical degradation like hydrolysis, oxidation, and other covalent bond formation or cleavage.\cite{castaneda2022alternative, kamerzell2011protein} While excipients have demonstrated the ability to stabilize proteins, the molecular mechanism(s) of stabilization for many protein-excipient combinations are not fully understood. Without a clear fundamental understanding of these stabilizing mechanisms, formulation development often relies on trial-and-error approaches, which can be time-consuming and expensive. \cite{dong2024formulationai, zarzar2023high} Elucidating the thermodynamic contributions of excipients to protein stability can facilitate the rational selection of stabilizers, enabling the design of formulations that withstand variations in storage and handling conditions.

Among the various classes of excipients, sugars are widely used in vaccine and protein formulations due to their demonstrated stabilizing effects, biocompatibility, and safety.\cite{castaneda2022alternative, shire2009formulation, singh2018sucrose, kamerzell_proteinexcipient_2011, bashir2020biophysical, chang2005mechanism, ajito2018sugar} In this study, we focus on the role of four sugars --  $\alpha$-glucose, $\beta$-fructose, trehalose, and sucrose -- in enhancing macromolecular stability. These sugars were selected due to their stabilizing effects in biological formulations\cite{castaneda2022alternative, shire2009formulation, singh2018sucrose, kamerzell_proteinexcipient_2011, bashir2020biophysical, chang2005mechanism, ajito2018sugar} and to investigate the role of molecular structure and size on macromolecular stabilization. Though there are several hypothesized mechanisms for sugar-induced macromolecular stability\cite{inoue_preferential_1972, scatchard_physical_1946, casassa_thermodynamic_1964, schellman_selective_1987, lin_role_1996, sudrik2019understanding, arsiccio2023thermodynamic, timasheff1998control, mensink2015line, chang2005mechanism, fahy2015principles, francia2008protein}, the extent to which these mechanisms can be leveraged for sugar excipient selection remains poorly understood. By investigating these hypothesized mechanisms, our study aims to provide insights that facilitate informed choices regarding sugar excipient selection in formulation development.

Proposed sugar-induced stabilization mechanisms include (a) preferential exclusion\cite{inoue_preferential_1972, scatchard_physical_1946, casassa_thermodynamic_1964, schellman_selective_1987, lin_role_1996, sudrik2019understanding}, (b) excluded volume\cite{arsiccio2023thermodynamic, timasheff1998control}, (c) water replacement\cite{mensink2015line}, (d) vitrification\cite{chang2005mechanism, fahy2015principles}, and (e) anchorage hypotheses\cite{francia2008protein}. While water replacement, vitrification, and anchorage primarily apply to freeze-dried formulations, preferential exclusion and excluded volume effects are particularly relevant to stabilization in liquid formulations. Liquid formulations are widely used and often preferred in biological applications due to their ease of processing, administration, and broad clinical applicability.\cite{bye2014biopharmaceutical, wang1999instability} This study aims to closely examine the hypothesized preferential exclusion and excluded volume mechanisms of sugar-induced stabilization in liquid formulations to guide the optimal selection of sugar excipients.

The preferential exclusion mechanism is rooted in thermodynamic principles that describe how cosolvents (like sugars) influence the surface free energy (the reversible work required to create a protein-sized cavity) in protein-solvent systems. \cite{timasheff1998control, kita_contribution_1994} When a protein is introduced into a solvent, it disrupts the hydrogen bonding network of water, creating an interface at the protein-solvent boundary. According to the Gibbs adsorption isotherm, extended to the context of protein-solvent systems, cosolvents that increase the air-water surface tension are preferentially excluded from the protein surface, leading to an enrichment of water molecules around the protein.\cite{timasheff1998control}  Several studies have observed weak or preferential exclusion of stabilizing sugars from the vicinity of the protein and have proposed the resulting preservation of the hydration shell as the reason for the observed protein stabilization. \cite{sudrik2019understanding, kim2018preferential, ajito2018sugar}

 This relationship between cosolvent-induced surface tension changes and preferential exclusion does not hold universally.\cite{kita_contribution_1994,barnett2016osmolyte, calero2017protein,wood2020hdx, olgenblum2023not} For example, cosolvents such as poly(ethylene glycol) (PEG), 2-methyl-2,4-pentanediol (MPD) and trimethylamine-N-oxide (TMAO) reduce the surface tension of water but are preferentially excluded from the protein, indicating that additional weak protein-cosolvent interactions play a role beyond surface tension effects.\cite{kita_contribution_1994} While preferential exclusion is widely observed for sugars, exceptions have been reported.\cite{barnett2016osmolyte, calero2017protein} For instance, sucrose has been found to preferentially accumulate around antistreptavidin immunoglobulin gamma-1 at high concentrations, whereas trehalose showed neither preferential interaction nor exclusion.\cite{barnett2016osmolyte} Both sucrose and trehalose have been observed to preferentially accumulate near $\alpha$-chymotrypsinogen, challenging the assumption that all sugars exhibit preferential exclusion behavior.\cite{calero2017protein} Recent studies suggest that preferential exclusion alone does not fully explain macromolecular stabilization, highlighting the role of specific sugar-protein interactions.\cite{wood2020hdx, olgenblum2023not} The excluded volume mechanism explicitly incorporates protein-excipient interactions into its thermodynamic framework. Cavity creation and protein-solvent interaction components are independently considered in the excluded volume mechanism, ultimately proposing that sugars stabilize proteins by favoring cavity creation in the native state.\cite{arsiccio2023thermodynamic, timasheff1998control}

Building on prior studies, our study assesses these mechanisms of sugar-induced protein stabilization and explores their implications for optimal sugar excipient selection. We perform molecular dynamics (MD) simulations at a level of detail and control that is difficult to achieve experimentally. These simulations enable a thermodynamic decomposition of interaction energies and entropic contributions that balance folding/unfolding. We utilize two polymer models -- a hydrophobic polymer (HP) and a charged polymer (CP) -- as simplified surrogates for proteins. These models capture key aspects of protein folding and unfolding while minimizing sequence complexity and computational cost, enabling a comprehensive exploration of sugar-mediated stabilization.\cite{dhabal2021characterizing, van2021length, athawale_enthalpyentropy_2008, athawale_osmolyte_2005, athawale_effects_2007} The HP model captures the influence of sugars on hydrophobic collapse, a dominant force in protein stability.\cite{dill_dominant_1990, ten2002hydrophobic, sun2022hydrophobic, pace2011contribution} The CP model builds on this basis to capture the role of electrostatic interactions in modifying sugar-mediated stabilization effects.

While this work was motivated by formulation design, the insights of this work are also relevant to understanding the influence of cosolvent environments on macromolecular folding and stability. Given that cytoplasm is a crowded environment comprising proteins, metabolites and osmolytes, it is important to understand protein folding equilibria in crowded environments.\cite{rivas2018toward, huber2025coil, cubukmacromolecular, monterroso2024macromolecular, dhar2010structure, christiansen2010factors} By quantifying the modulations of polymer conformations from sugar-induced compositional and entropic effects, our framework offers a simplified molecular perspective relevant to crowding phenomena in cellular environments.

 Our results reveal that sugar-induced polymer stabilization follows a concentration-dependent trend, with stabilization observed at low sugar concentrations and a transition toward unfolding at higher concentrations. A cut-off value for the preferential interaction coefficient ($\Gamma$) is identified, beyond which sugars begin to favor unfolding. Thermodynamic decomposition highlights the competing roles of polymer-sugar interactions, polymer-water interactions, and entropic effects in determining stability, with entropic contributions consistently promoting unfolding across all sugar solutions. Additionally, using a local mixing entropy metric, we demonstrate that polymer-solvent interaction entropy plays a key role in polymer stability at high sugar concentrations. The insights gained from these molecular-level analyses contribute towards rational selection of sugars for stabilizing proteins in liquid formulations.

\section{Methods}
\subsection{System Setup and Molecular Dynamics Simulations}

\begin{figure}[!ht]
 \centering
 \includegraphics[width=1\textwidth]{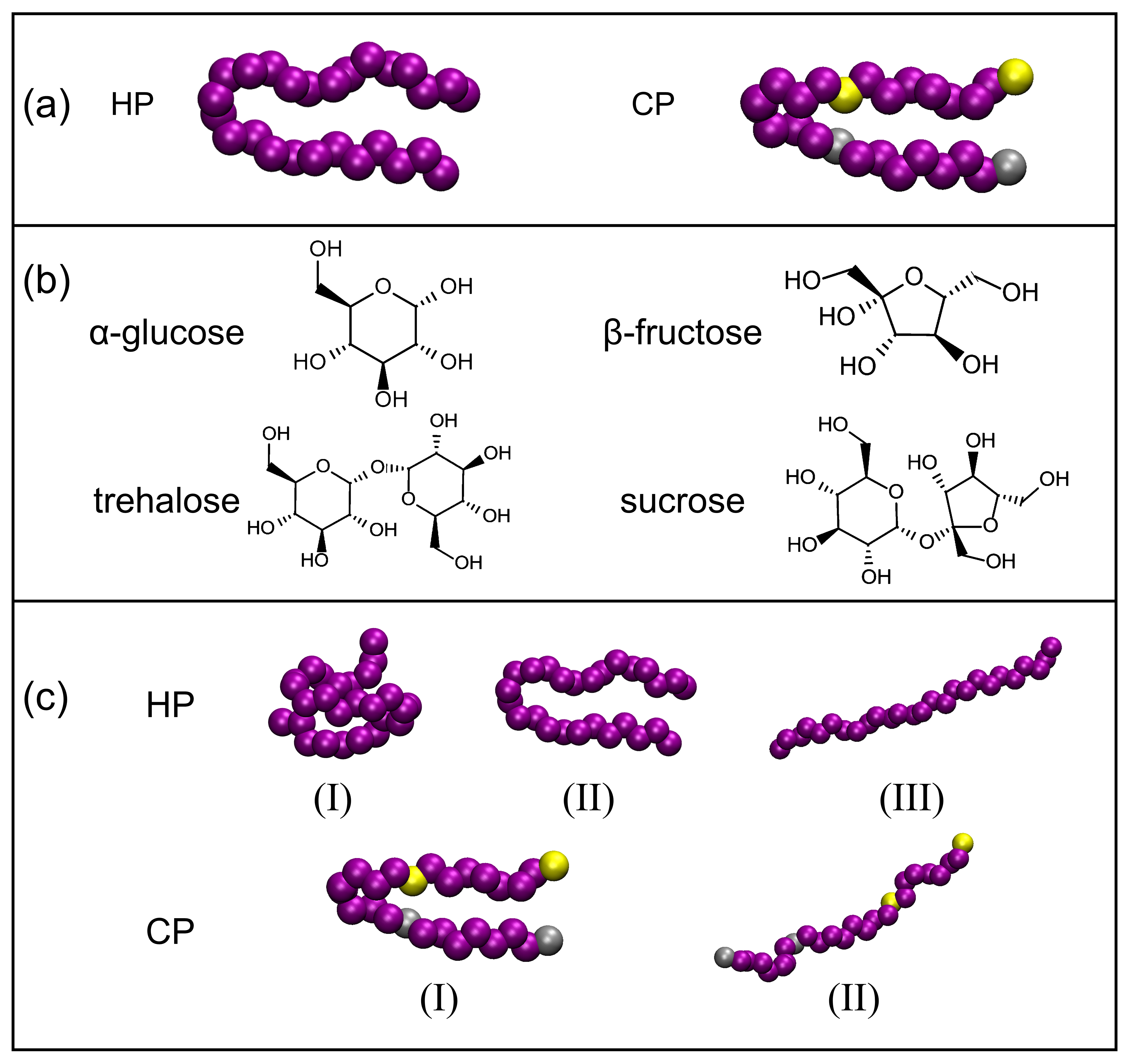}
 \caption{ (a) The polymer models considered in the current study. The colors of the beads correspond to their partial charges: purple $\rightarrow$ 0, silver $\rightarrow$ +0.5, yellow $\rightarrow$ -0.5, (b) 2D structures of the sugars considered in this study, (c) Representative polymer conformations: Top --  fully collapsed (I), hairpin (II), and unfolded state (III) for HP; Bottom --  hairpin (I) and unfolded state (II) for CP.}
    \label{fig:molecules}
\end{figure}

We simulated a united-atom HP model with 26 monomeric units, where each monomer represents a methylene ($\mathrm{CH_2}$) unit in terms of size and mass. The Lennard-Jones parameters used for the polymer model are the same as those used previously to represent $\mathrm{CH_2}$.\cite{zajac2025flipping,zajac2025impact, athawale_osmolyte_2005} ($\sigma= 0.373$ $\mathrm{nm}$ and $\epsilon= 0.5856$ $\mathrm{
 kJ/mol}$). The CP model is similar to the HP model with additional partial charges ($\pm 0.5$ e) assigned to 4 beads (Fig. \ref{fig:molecules}a). The charges were arranged in a (++––) sequence to mimic localized charged patches on protein surfaces, which promote intrachain electrostatic attraction and a hairpin-like folded configuration. The net charge of the CP model is zero, ensuring that any observed differences arise from local electrostatic effects rather than overall charge imbalance. Replica exchange umbrella sampling (REUS) simulations (see Section \ref{sec-reus}) were performed using GROMACS 2021.7 with the PLUMED 2.9.0 patch applied. Kirkwood-Buff derived parameters (KBP) developed by Cloutier et al. \cite{cloutier2018kirkwood} were used for the sugars ($\alpha$-glucose, $\beta$-fructose, trehalose, sucrose) considered in this study (Fig. \ref{fig:molecules}b). The KBP models employ an all-atom representation of carbohydrates and more accurately describe protein-excipient-water interactions than standard force fields.

We employed the TIP4P/2005 water model\cite{abascal2005general} to describe water. TIP4P/2005 has been shown to accurately predict several water properties such as density, viscosity, diffusion coefficients, and the hydrogen-bonding network \cite{vega2007surface,abascal2005general}. Since the original KBP were developed for use with the TIP3P water model,\cite{jorgensen1983comparison} we assessed their applicability to the TIP4P/2005 water model. We calculated the Kirkwood-Buff integral for the sugar-water systems and compared the results with those reported by Cloutier et al.\cite{cloutier2018kirkwood}. The values obtained are within the range of experimental error, confirming the compatibility of KBP with TIP4P/2005 model (see Table S1 for details). We performed MD simulations of HP and CP models in sugar solutions of concentrations ranging from 0 to 2.0 M (see Table S2 for further details). Monosaccharide concentrations were studied up to 2.0 M and disaccharide concentrations were limited to 1.0 M to ensure an equivalent monomeric unit concentration.

 The systems in all simulations were first energy minimized using the steepest descent algorithm.\cite{van_der_spoel_gromacs_2005} NVT equilibration was carried out for 1 ns at 300 K, followed by a 1 ns NPT equilibration at 300 K and 1 atm. During equilibration, temperature was controlled with the V-rescale thermostat,\cite{bussi_canonical_2007} and pressure was controlled with the Berendsen barostat.\cite{berendsen_molecular_1984}  Following equilibration, NPT production runs were completed using the Nos\'e-Hoover thermostat\cite{evans_nosehoover_1985} and Parrinello-Rahman barostat.\cite{parrinello_polymorphic_1981} Production runs were completed for 100 ns per window for polymer/sugar/water REUS simulations. 100 ns was sufficient to achieve convergence, as demonstrated by the evolution of the PMF with simulation time (SI Fig. S1). The Particle Mesh Ewald (PME) algorithm was used for long-range electrostatic interactions with a cut-off of 1 $\mathrm{nm}$.\cite{darden1993particle} A reciprocal grid of 42 x 42 x 42 cells was used with $\mathrm{4^{th}}$ order B-spline interpolation. A single cut-off of 1 $\mathrm{nm}$ was used for van der Waals interactions. The neighbor search was performed every 10 steps. Lorentz-Berthelot mixing rules\cite{lorentz_1881, berthelot1898melange} were used to calculate non-bonded interactions between different atom types, except for polymer-water oxygen interactions. The polymer-water oxygen interactions were adjusted such that the folded and unfolded states of the polymer in pure water were approximately evenly distributed in straightforward MD simulations.\cite{zajac2025flipping, zajac2025impact} 

\subsection{Replica Exchange Umbrella Sampling} \label{sec-reus}

REUS was employed to sample the conformational free energy landscape of the polymer in different sugar solutions and obtain the potential of mean force (PMF) as a function of the polymer radius of gyration ($R_g$). 12 evenly spaced umbrella potential window centers from 0.35 to 0.90 nm were used, with a spacing of 0.05 nm between consecutive centers. A harmonic bias potential of 5000 kJ/mol/nm$^2$ was applied to all the windows except the window at 0.45 nm, where 1000 kJ/mol/nm$^2$ was applied. This choice of force constants was adopted from our previous work on a similar system, where it ensured adequate sampling and convergence of the free energy profile.\cite{zajac2025flipping, zajac2025impact}

The free energy of polymer unfolding ($\Delta G_u = G_{unfolded} - G_{folded}$) was calculated according to:

\begin{equation}\label{eqn:dgu}
     {\Delta G_{\text{u}}=-k_{B}T\ln\left({\frac{\int_{R_{g,cut }}^{R_{g,\max}} \exp\left(\frac{-W\left(R_g\right)}{k_B T}\right) dR_g}{\int_{R_{g,\min}}^{R_{g,cut }} \exp\left(\frac{-W\left(R_g\right)}{k_B T}\right) dR_g}}\right)}
\end{equation}

\noindent where $R_{g,cut}$ is determined as the transition point between the folded and unfolded states where $\frac{ dW(R_{g})}{dR_{g}} = 0$.  $W(R_g)$ is the PMF, calculated as ${W(R_g) =}$ $ {-k_{B}T \ln(P(R_g))}$. The Weighted Histogram Analysis Method (WHAM) was used to reweight biased probability distributions obtained from REUS simulations.\cite{zhu_convergence_2012}

\subsection{Thermodynamic Decomposition} \label{sec-decomp}

We decomposed the PMF into thermodynamic components to further investigate the role of sugars in HP and CP folding. Following the methods established in prior studies,\cite{athawale_effects_2007, godawat_unfolding_2010,dasetty_advancing_2021} the PMF can be separated into two contributions -- intrapolymer interactions and solvent-mediated interactions ($W_{solv}$). The former can be estimated from simulations of polymer in vacuum ($W_{vac}$), leading to the expression:

\begin{equation}
    W(R_g) = W_{vac}(R_g) + W_{solv}(R_g)
\label{eqn:decompA}
\end{equation}

\noindent Here, both $W(R_g)$ and $W_{vac}(R_g)$ are obtained directly from simulations: $W(R_g)$ is the potential of mean force (PMF) for the polymer in solution, and $W_{vac}(R_g)$ is computed for the same polymer in vacuum. The solvation contribution, $W_{solv}(R_g)$, is then evaluated from Eq.~\ref{eqn:decompA} as their difference.

Considering a two-state model for polymer folding, the free energy of unfolding in any solvent can be written as:

\begin{equation}
    \Delta G_u = \Delta G_{vac} + \Delta G_{solv}
\label{eqn:decompB}
\end{equation}

Consistent with the solvation theory approaches, \cite{pratt_theory_1977, lum_hydrophobicity_1999, wolde_hydrophobic_2002, pratt_hydrophobic_2002, huang_hydrophobic_2002,chandler_interfaces_2005, garde_temperature_1999, ben-amotz_water-mediated_2016, arsiccio2023thermodynamic, schellman2003protein} $\Delta G_{solv}$ can be conceptually decomposed into two steps: (1) the reversible work required to create a polymer-sized cavity in the solvent, and (2) the energy contributions arising from "turning on" polymer-solvent interactions upon solvation. This relationship was expressed by Arsiccio et al.\cite{arsiccio2023thermodynamic} as: 

\begin{equation}
    \Delta G_{solv} = \Delta G_{ex} + \Delta G_{si}
\label{eqn:decompC}
\end{equation}
\noindent where: \begin{equation}
           \Delta G_{ex} = \Delta H_{ex} - T\Delta S_{ex},\ 
           \Delta G_{si} = \Delta H_{si} - T\Delta S_{si}
\label{eqn:decompD}
\end{equation}

\noindent Here, $\Delta G_{ex}$ represents the free energy associated with creating a cavity in the solvent (also known as the excluded volume contribution), and $\Delta G_{si}$ represents the contribution from polymer-solvent interactions (also known as the soft-interaction contribution). The cavity formation step primarily reflects the energetic cost of disrupting solvent–solvent interactions and increasing the solvent surface area around the polymer. The small PV-work associated with the change in system volume is included implicitly, but it is negligible at ambient pressure, making the interfacial (surface-tension) contribution dominant.

The total enthalpic contribution ($\Delta H_{ex}$ + $\Delta H_{si}$) can therefore be approximated as the polymer–solvent interaction energy, assuming negligible PV-work:

\begin{equation}
           \Delta H_{ex} + \Delta H_{si} \approx \Delta E_{p-solv} =  
           \Delta E_{pw} + \Delta E_{ps}
\label{eqn:decompE}
\end{equation}

\noindent where $\Delta E_{pw}$ and $\Delta E_{ps}$ represent the changes in ensemble-averaged interaction energies between the folded and unfolded states of the polymer. Specifically, $\Delta E_{pw}$ corresponds to the polymer-water interaction energy, while $\Delta E_{ps}$ corresponds to the polymer-sugar interaction energy. Eq. \ref{eqn:decompC} can be rewritten as:

\begin{equation}
           \Delta G_{solv} = \Delta E_{p-solv} - T\Delta S_{ex} - T\Delta S_{si}\
            = \Delta E_{p-solv} + \Delta G_{rem}
\label{eqn:decompF}
\end{equation}

$\Delta G_{u}$ can thus be written as the following  by combining Eq. \ref{eqn:decompB} and Eq. \ref{eqn:decompF}: 
\begin{equation}
\Delta G_{u} =  \Delta G_{vac} + \Delta E_{p-solv} +  \Delta G_{rem}
\label{eqn:decompG}
\end{equation}
This decomposition framework closely resembles the methodology employed by Athawale et al. \cite{athawale_effects_2007} for a hydrophobic polymer in water, where the $\langle U_{pw} \rangle$ and $\Delta G_{hyd}$ terms in their work are analogous to $E_{p-solv}$ and $\Delta G_{rem}$ in our formulation.

In Eq \ref{eqn:decompG}., the interaction energies $\Delta E_{pw}$ and $\Delta E_{ps}$ were computed using the \textit{gmx rerun} functionality in GROMACS, with separate energy groups defined for polymer, sugar, and water. $\Delta G_{u}$ is calculated using Eq. \ref{eqn:dgu}. $\Delta G_{rem}$ can then be determined by rewriting Eq. \ref{eqn:decompG} as $\Delta G_{rem}$ = $\Delta G_{u} - \Delta G_{vac} - \Delta E_{p-solv}$. 

The change in free energy of unfolding relative to pure water upon sugar addition is given by:
\vspace{-0.1cm}
\begin{equation}
\begin{aligned}
   \Delta\Delta G_u & = \Delta G_{u,sugar-soln} - \Delta G_{u,water} \\ & = \Delta\Delta G_{vac} +  \Delta\Delta E_{p-solv} + \Delta\Delta G_{rem} 
    \\ & = \Delta\Delta E_{pw} + \Delta\Delta E_{ps} + \Delta\Delta G_{rem}
\label{eqn:decompH}
\end{aligned}
\end{equation}

Since the vacuum contribution has negligible dependence on the solvent, it cancels out in Eq. \ref{eqn:decompH}. Positive values of $ \Delta\Delta G_u$ correspond to the stabilization of the folded state by the sugar, and negative values correspond to the destabilization of the folded state.

In summary, $W(R_g)$ and $W_{vac}(R_g)$ were obtained directly from simulations of the polymer in solution and vacuum, $W_{solv}(R_g)$ from their difference, $\Delta E_{pw}$ and $\Delta E_{ps}$ from energy decomposition via \texttt{gmx rerun}, and $\Delta G_{rem}$ by difference using Eq.~\ref{eqn:decompG}. This approach cleanly isolates solvent-mediated contributions governing sugar-induced polymer stabilization.

\subsection{Preferential Interaction Coefficient} \label{section-pref-int}

The distribution of sugar and water molecules around the polymer can be described via the preferential interaction coefficient ($\Gamma_{ps}$),\cite{scatchard_physical_1946, casassa_thermodynamic_1964, schellman_selective_1987}
\begin{equation}
    \Gamma_{ps}=-\left(\frac{\partial \mu_p}{\partial \mu_s}\right)_{m_p, T, P}=\left(\frac{\partial m_s}{\partial m_p}\right)_{\mu_s, T, P}
\end{equation}
where $\mu$ is the chemical potential, $m$ is the concentration in molality ($\mathrm{mol/kg}$), subscripts  $p$, and $s$ refer to polymer and sugar, respectively, and T and P refer to the temperature and pressure. This parameter is calculated in simulations using the two-domain formula\cite{inoue_preferential_1972, record_interpretation_1995, shukla_molecular_2009} given by:
\begin{equation}
    \Gamma_{ps}=\left\langle N_s^{\text {local }}-\left(\frac{N_s^{\text {bulk }}}{N_w^{\text {bulk }}}\right) N_w^{\text {local }}\right\rangle
\end{equation}
\noindent
where $N$ represents the number of molecules of a given species, $w$ represents water, and angular brackets denote an ensemble average. The local and bulk domain was separated by a cut-off distance $R_{cut}=1.2$ nm from the polymer, as $\Gamma_{ps}$ values had minimal variation beyond this distance (SI Figs. S2 and S3). $\Gamma_{ps}$ gives a measure of the relative accumulation or depletion of sugar in the local domain of the polymer with respect to the bulk domain, with $\Gamma_{ps} > 0$ indicating relative accumulation (preferential interaction) and $\Gamma_{ps} < 0$ indicating relative depletion (preferential exclusion).

\subsection{Local Mixing Entropy} \label{section-local-mixing-entropy}

To quantify the heterogeneity of the local solvent environment surrounding the polymer, we compute the local mixing entropy using atomic number fractions of water and sugar molecules within a radial shell around the polymer. While the classical expression for entropy of mixing is based on mole fractions of molecular species, here we adopt a modified approach using atom number fractions to reflect the fine-grained spatial distribution and diversity of interactions among solvent components near the polymer. 
The local mixing entropy is computed as:
\begin{equation}
    \frac{\Delta S_{mix}}{R}=\left\langle \sum_{i={s, w}} -x_i \ln x_i \right\rangle
\end{equation}
where $x_i$ corresponds to the atom number fraction of component $i$ in the polymer's solvation shell, with $i$ representing either sugar ($s$) or water ($w$), and angular brackets denote an ensemble average. These fractions are calculated as:
\begin{equation}
    x_i=\frac{n_i}{n_{s} + n_{w}}
\end{equation}
with $n_i$ denoting the number of atoms of component $i$ found within the local environment of the polymer for a given timeframe.
Although this formulation departs from the conventional mole-fraction-based entropy of mixing, it is used here as a relative metric to compare the diversity and configurational complexity of the solvent environment in the vicinity of the polymer across different conformational states and sugar compositions. The underlying rationale is that a higher atom-based local mixing entropy suggests a more heterogeneous solvation environment, potentially enabling a wider range of polymer–solvent interactions. To assess how solvent heterogeneity changes upon polymer unfolding, we compute the difference in local mixing entropy between the folded and unfolded states:

\begin{equation}
    \Delta \Delta S_{mix}= \Delta S_{mix, folded} - \Delta S_{mix, unfolded}
\end{equation}

A positive value of $\Delta \Delta S_{mix}$ indicates that the folded state is associated with a more diverse solvent environment. Such diversity reflects a broader ensemble of accessible polymer–solvent configurations, suggesting that the folded state permits greater variability in how solvent molecules are arranged around the polymer. The resulting configurational diversity increases system entropy and may enhance the thermodynamic stability of the folded state. Overall, this metric provides insight into the entropic modulation of solvent–polymer interactions and complements the energetic analyses in understanding sugar-mediated stabilization mechanisms.

\section{Results and Discussion}

This study aims to investigate the effect of sugars on the thermodynamics of polymer unfolding in aqueous solutions, with implications for understanding protein stabilization in formulations. Specifically, we examine the role of two isomeric monosaccharides ($\alpha$-glucose, $\beta$-fructose), their mixture ($\alpha$-glucose + $\beta$-fructose), and two disaccharides (trehalose, sucrose) on the stability of HP and CP models. By analyzing the polymer unfolding free energies, preferential interactions, entropic contributions, and energetic effects in these systems, we aim to elucidate the role of different sugars in protein stabilization mechanisms.

\subsection{Effect of Sugars on Polymer Folding}

\begin{figure*}[!ht]
 \centering
 \includegraphics[width=1\textwidth]{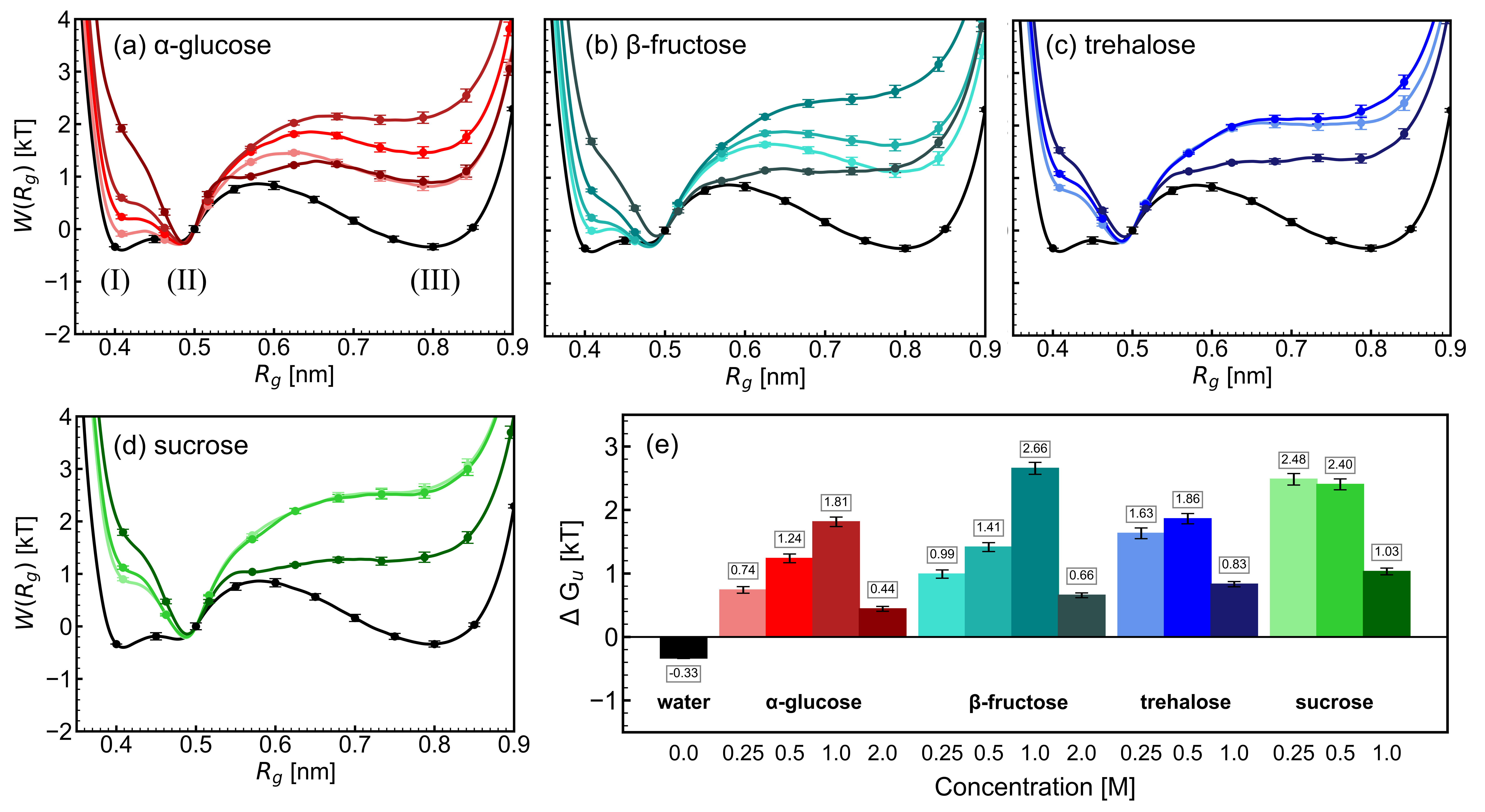}
 \caption{Hydrophobic polymer PMFs and free energy of hydrophobic polymer unfolding in different excipient solutions. Red -- $\alpha$-glucose, cyan -- $\beta$-fructose, blue -- trehalose, green -- sucrose. Increasing shading corresponds to increasing concentration of sugars. Increasing additive concentration is denoted by increased shading (light to dark). Error bars were estimated from bootstrapping (100 samples).}
    \label{fig:hp-pmf}
\end{figure*}

The PMF reveals key stable states and the energetic barriers between them. Figs. \ref{fig:hp-pmf}a-\ref{fig:hp-pmf}d show the PMF for the HP across the sugar solutions under study. Stable conformational states are observed at (I) $R_g \sim 0.4$ nm (fully collapsed), (II) $R_g \sim 0.5$ nm (hairpin), and (III) $R_g \sim 0.8$ nm (unfolded), with states (I) and (II) belonging to the folded ensemble. Representative polymer conformations for these states are shown in Fig. \ref{fig:molecules}c. Among these states, the hairpin state around $R_g \sim 0.5$ nm consistently emerged as the most stable configuration across all solutions, making it the reference state for comparing PMFs.

\begin{figure*}[!ht]
 \centering
 \includegraphics[width=1\textwidth]{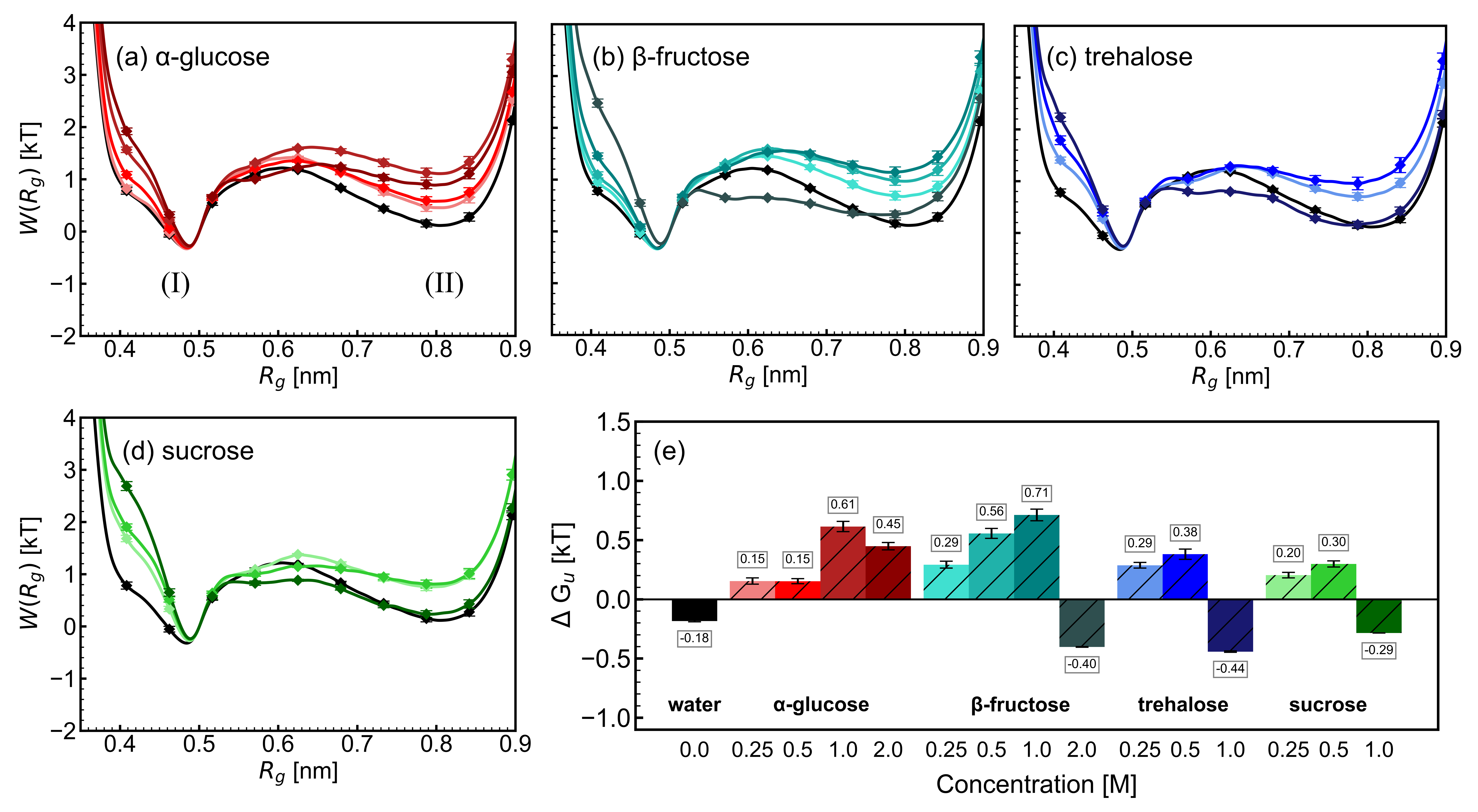}
 \caption{Charged polymer PMFs and free energy of charged polymer unfolding in different excipient solutions. Red -- $\alpha$-glucose, cyan -- $\beta$-fructose, blue -- trehalose, green -- sucrose. Increasing shading corresponds to increasing concentration of sugars. Increasing additive concentration is denoted by increased shading (light to dark). Error bars were estimated from bootstrapping (100 samples). Hatching patterns are used to differentiate CP results from HP.}
    \label{fig:cp-pmf}
\end{figure*}

Fig. \ref{fig:hp-pmf}a-\ref{fig:hp-pmf}d shows that the stability of the fully collapsed state ($R_g \sim 0.4$ nm) decreases with increasing sugar concentration, suggesting the destabilization of compact conformations in more concentrated sugar solutions. The unfolded state ($R_g \sim 0.8$ nm), on the other hand, displayed distinct trends depending on the sugar type. In monosaccharide solutions ($\alpha$-glucose, $\beta$-fructose), the stability of the unfolded state decreased with increasing sugar concentrations up to 1 M, followed by a partial stabilization at 2 M. In disaccharide solutions (trehalose, sucrose), the unfolded state initially destabilized up to 0.5 M, followed by a partial stabilization at 1 M. Notably, the onset of unfolded state stabilization occurs at roughly half the concentration in disaccharide solutions compared to monosaccharide solutions. This pattern suggests a potential size-dependent mechanism, where the larger disaccharides reach the inflection point at lower concentrations. A more detailed analysis of these effects, including their enthalpic and entropic contributions, is explored in Section \ref{sec-pmf-decomp}.

The free energy of unfolding provides a measure of the stability of the ensemble of the folded vs. unfolded states of the polymer. For monosaccharides, the free energy of unfolding increases as the sugar concentration increases up to 1 M. This is consistent with previous observations using the linear extrapolation method, which assumes that the free energy of unfolding varies monotonically with cosolvent concentrations and is commonly used to estimate stability changes in the presence of denaturants or stabilizers. \cite{canchi_cosolvent_2013, pace1975stability, arsiccio2023thermodynamic} However, at 2 M concentration, the free energy of unfolding decreases, suggesting a non-monotonic, concentration-dependent influence of sugars on polymer stability. This shift occurred at lower concentrations for disaccharides (0.5 M) than monosaccharides, consistent with the trends observed in the unfolded state region of the PMF. 

In the case of the CP, the partial charges introduce additional electrostatic interactions that influence the stability of the polymer's conformational states, and its interactions with the solvent. Fig. \ref{fig:cp-pmf}a-\ref{fig:cp-pmf}d depicts the PMF for the CP. For CP, stable states were observed at $R_g \sim 0.5$ nm (hairpin) and $R_g \sim 0.8$ nm (unfolded). Representative conformations for these CP states are shown in Fig. \ref{fig:molecules}c (bottom). A key difference between HP and CP is the relative instability of the fully collapsed state ($R_g \sim 0.4$ nm) in CP compared to HP. This relative instability is likely due to a combination of: (i) attractive polymer-solvent charge interactions in the hairpin and unfolded states, and (ii) like-charge repulsion effects in the fully folded state. 

Upon adding sugar, the fully collapsed state ($R_g \sim 0.4$ nm) for the CP becomes less stable with increasing concentrations (Fig. \ref{fig:cp-pmf}a-\ref{fig:cp-pmf}d). The unfolded state ($R_g \sim 0.8$ nm) initially destabilizes, followed by partial stabilization at high concentration, similar to the trends observed for the HP. The stabilizing effect of sugars at low concentrations was weaker in CP than in HP. This weaker stabilizing effect, combined with the stabilization of the unfolded state at high concentrations, resulted in an overall preference for the unfolded state at 2 M $\beta$-fructose, 1 M trehalose, and 1 M sucrose. For 2 M $\alpha$-glucose, similar trends are observed, but the magnitude of these effects was insufficient to favor the unfolded state overall.

Having established the effects of sugars on the stability of different states in both the HP and CP models, we now focus on uncovering the underlying mechanisms responsible for these observations. Specifically, we evaluate the extent to which the proposed mechanisms of sugar-induced macromolecule stabilization, as discussed in the literature, \cite{inoue_preferential_1972, scatchard_physical_1946, casassa_thermodynamic_1964, schellman_selective_1987, lin_role_1996, sudrik2019understanding, arsiccio2023thermodynamic, timasheff1998control, mensink2015line, chang2005mechanism, fahy2015principles, francia2008protein} account for the observations in our systems. In the following section, we focus on the preferential exclusion mechanism.

\subsection{Role of Preferential Exclusion on Polymer Stabilization} \label{sec-pref-exc}

\begin{figure}[!ht]
 \centering
 \includegraphics[width=1\textwidth]{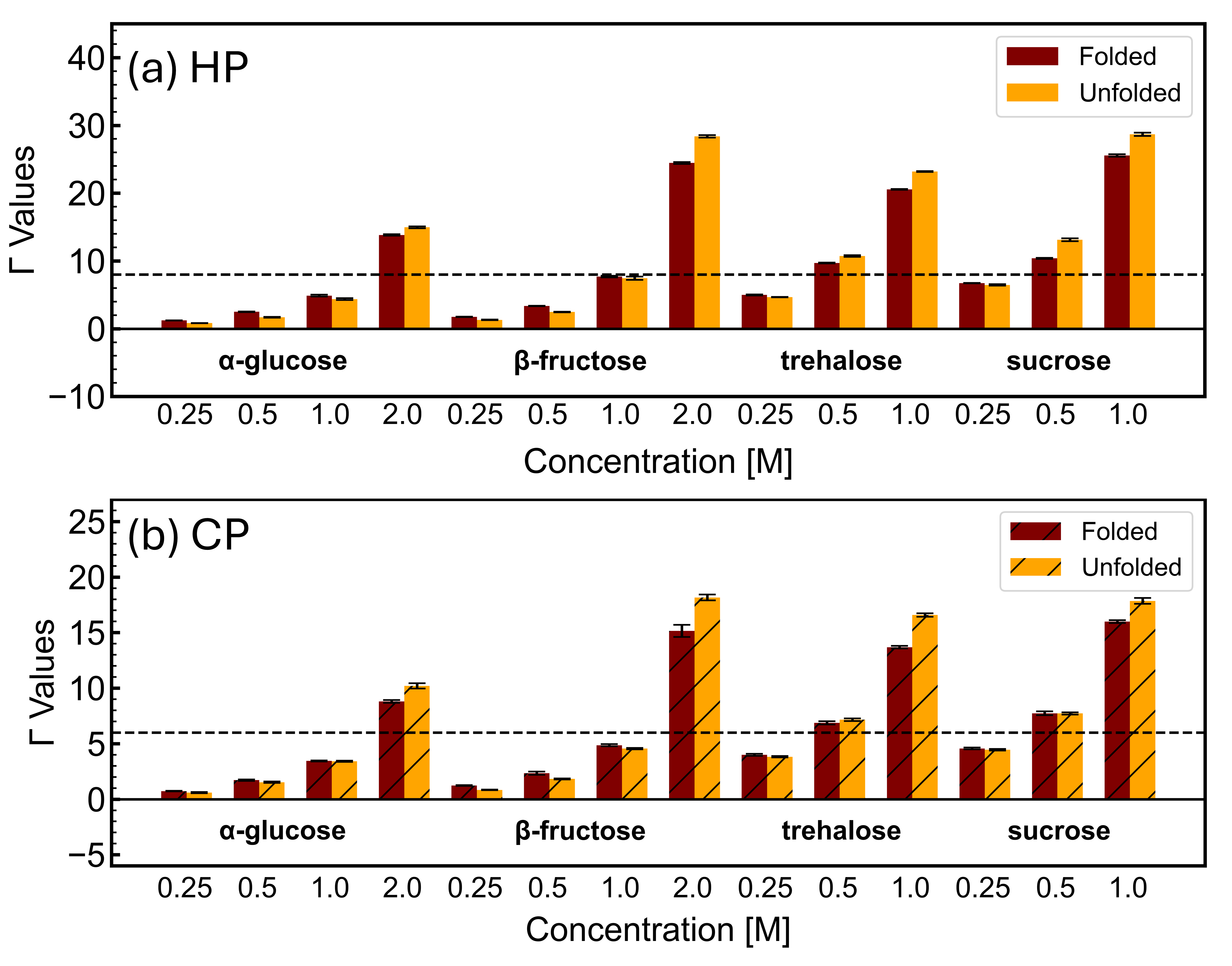}
 \caption{The preferential interaction coefficients for the folded and unfolded state of (a) HP and (b) CP in different sugar solutions. The cut-off separating the local and bulk domains is set to 1.2 nm. Refer to SI Figs. S2 and S3 for the variation of $\Gamma$ with the choice of this cut-off. Mean $\Gamma_{ps}$ of the last four 20 ns simulation trajectory blocks and the standard error of those means are reported.}
    \label{fig:pref-int}
\end{figure}

We compute the preferential interaction coefficient ($\Gamma$) for the HP and CP models in different sugar solutions. Our results consistently show positive $\Gamma$ values across all sugar solutions, indicating preferential accumulation of sugars around HP and CP. The magnitude of $\Gamma$ increases with sugar concentration, with disaccharides exhibiting higher values than monosaccharides at the same molar concentrations. To further assess whether these trends arise purely from molecular concentration or from the total sugar content, we also compared $\Gamma$ values on a mass basis, that is, for solutions containing approximately equivalent total sugar mass (e.g., 1 M trehalose vs. 0.5 M glucose). Even under these mass-equivalent conditions, disaccharides continue to exhibit higher $\Gamma$ values than monosaccharides, with the notable exception of 2 M $\beta$-fructose, which shows values comparable to 1 M sucrose. This comparison is relevant for formulation design, where excipient levels are often specified by weight percent rather than molar concentration

% On a mass basis, disaccharides still tend to exhibit higher $\Gamma$ values than monosaccharides, with the notable exception of 2 M $\beta$-fructose, which shows values comparable to 1 M sucrose.

The preferential exclusion mechanism \cite{timasheff1998control, arsiccio2023thermodynamic} suggests that stabilizing sugars are preferentially excluded from (folded) proteins, resulting in negative $\Gamma$ values. However, our results show positive $\Gamma$ values even in cases where the folded states of HP and CP are thermodynamically favored, seemingly contradicting this mechanism. This apparent inconsistency can be reconciled by recognizing that preferential exclusion is fundamentally governed not by the absolute magnitude of $\Gamma$, but by the difference in $\Gamma$ between the folded and unfolded states. Specifically, the mechanism assumes greater exclusion from the unfolded state, and it is this difference in $\Gamma$ that contributes to the stabilization. To capture this effect, we examine the difference in preferential interactions: $\Delta \Gamma = \Gamma_{folded} - \Gamma_{unfolded}$. At lower sugar concentrations, $\Delta \Gamma$ is positive, indicating stronger sugar accumulation around the folded state. This correlates with an increase in unfolding free energy with concentration, suggesting increasing stabilization of the folded state. At higher concentrations, $\Delta \Gamma$ becomes negative, and this transition coincides with a decrease in unfolding free energy with increasing sugar concentration. This trend is consistent with the Wyman linkage relation \cite{wyman_linked_1964}, which links changes in solute binding to changes in macromolecular stability. Wyman linkage relation for folding/unfolding equilibria can be expressed as:\cite{wyman_linked_1964}

\begin{equation}
    \frac {\partial \Delta G_{u}}{\partial c_{s}} = \frac{\partial \mu_{s}}{\partial (\text{ln} c_{s})} \Delta \Gamma
\end{equation} 

According to the preferential exclusion mechanism, this transition of $\Delta \Gamma$ from positive to negative is expected to occur at $\Gamma = 0$ for proteins. In contrast, we observe that this crossover occurs at a nonzero $\Gamma$, which we denote as $\Gamma_{cut}$. We heuristically identify $\Gamma_{cut} \approx 8$ for HP and $\Gamma_{cut} \approx 6$ for CP (marked by dashed lines in Fig. \ref{fig:pref-int}a, b) as the value that separates positive from negative $\Delta \Gamma$ values. Interestingly, we also find that $\Gamma$ values are lower for CP compared to HP. This suggests that the presence of partial charges on the polymer reduces sugar accumulation and provides a possible explanation for why proteins, which contain many partial charges, often exhibit preferential exclusion.\cite{sudrik2019understanding, ajito2018sugar, arsiccio2023thermodynamic, timasheff1998control, kita_contribution_1994, kim2018preferential} The relationship between changes in $\Delta G$ and $\Delta \Gamma$ is discussed in SI Fig. S4.

While preferential interaction coefficients offer qualitative insight into concentration-dependent stability trends, they do not quantitatively predict whether folding or unfolding is thermodynamically favored at any given concentration. For instance, in HP with 2 M $\beta$-fructose, the unfolded state exhibits a higher $\Gamma$ than the folded state, suggesting that $\frac{\partial \Delta G_u}{\partial c_{s}} < 0$. However, 2 M $\beta$-fructose still stabilizes the folded state relative to pure water ($\Delta G_{u,\, \text{2 M }\beta\text{-fructose}} - \Delta G_{u, water} > 0$). This underscores the need to go beyond preferential interaction values and examine the underlying thermodynamic contributions that govern folding and unfolding. To address this, we perform a thermodynamic decomposition of the free energy of unfolding in the following section (see Section \ref{sec-pmf-decomp}), which provides a more comprehensive view of the interactions and entropic factors governing polymer stability.

\subsection{Thermodynamic Decomposition of Unfolding Free Energy}  \label{sec-pmf-decomp}

\begin{figure*}[!ht]
 \centering
 \includegraphics[width=1\textwidth]{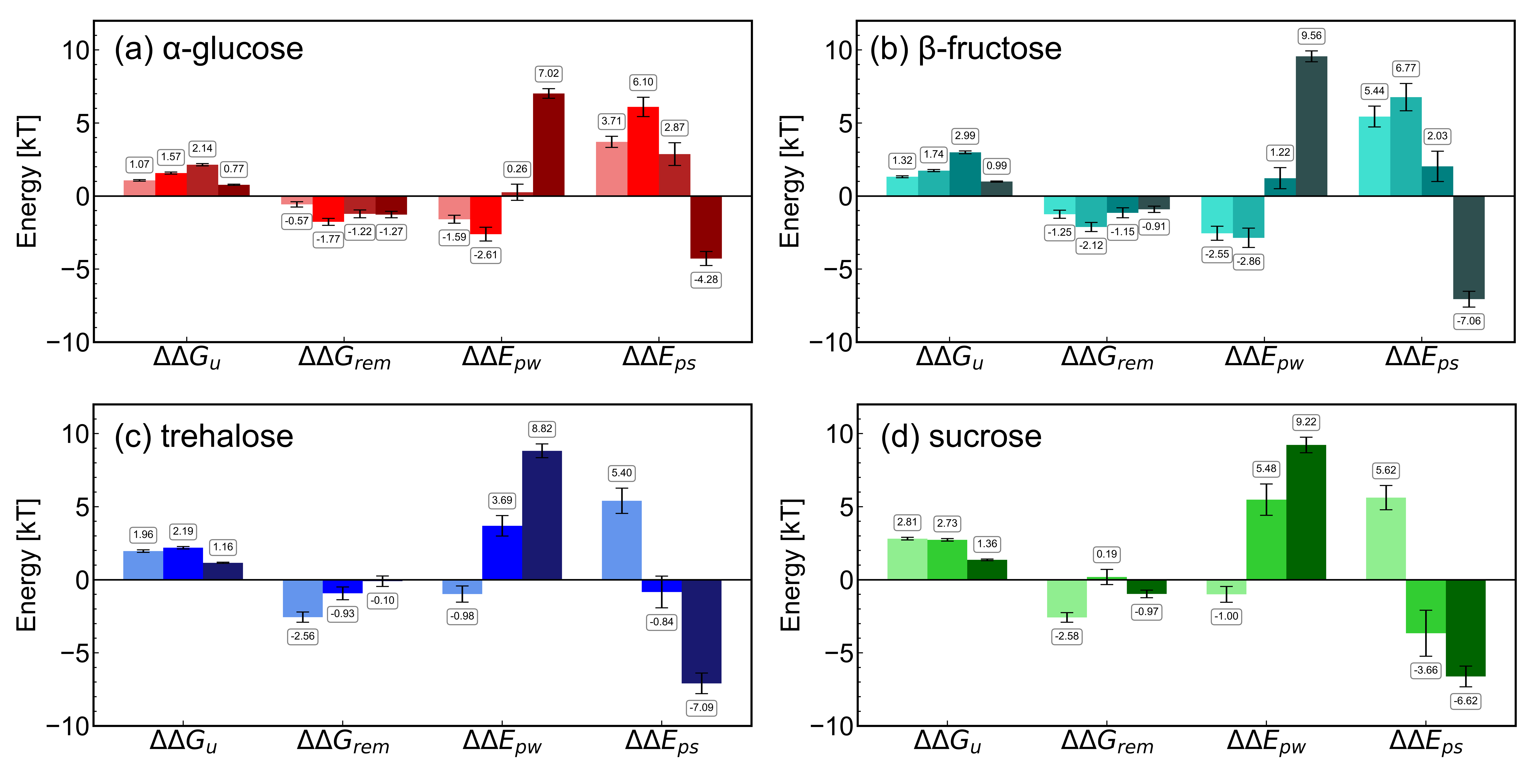}
 \caption{Decomposition of HP unfolding free energy contributions in (a) $\alpha$-glucose, (b) $\beta$-fructose, (c) trehalose, and (d) sucrose solutions. Positive values make the unfolding in the presence of sugar molecules less favorable than in pure water. Increasing additive concentration is denoted by increased shading (light to dark).}
    \label{fig:hp-pmf-decomp}
\end{figure*}

\begin{figure*}[!ht]
 \centering
 \includegraphics[width=1\textwidth]{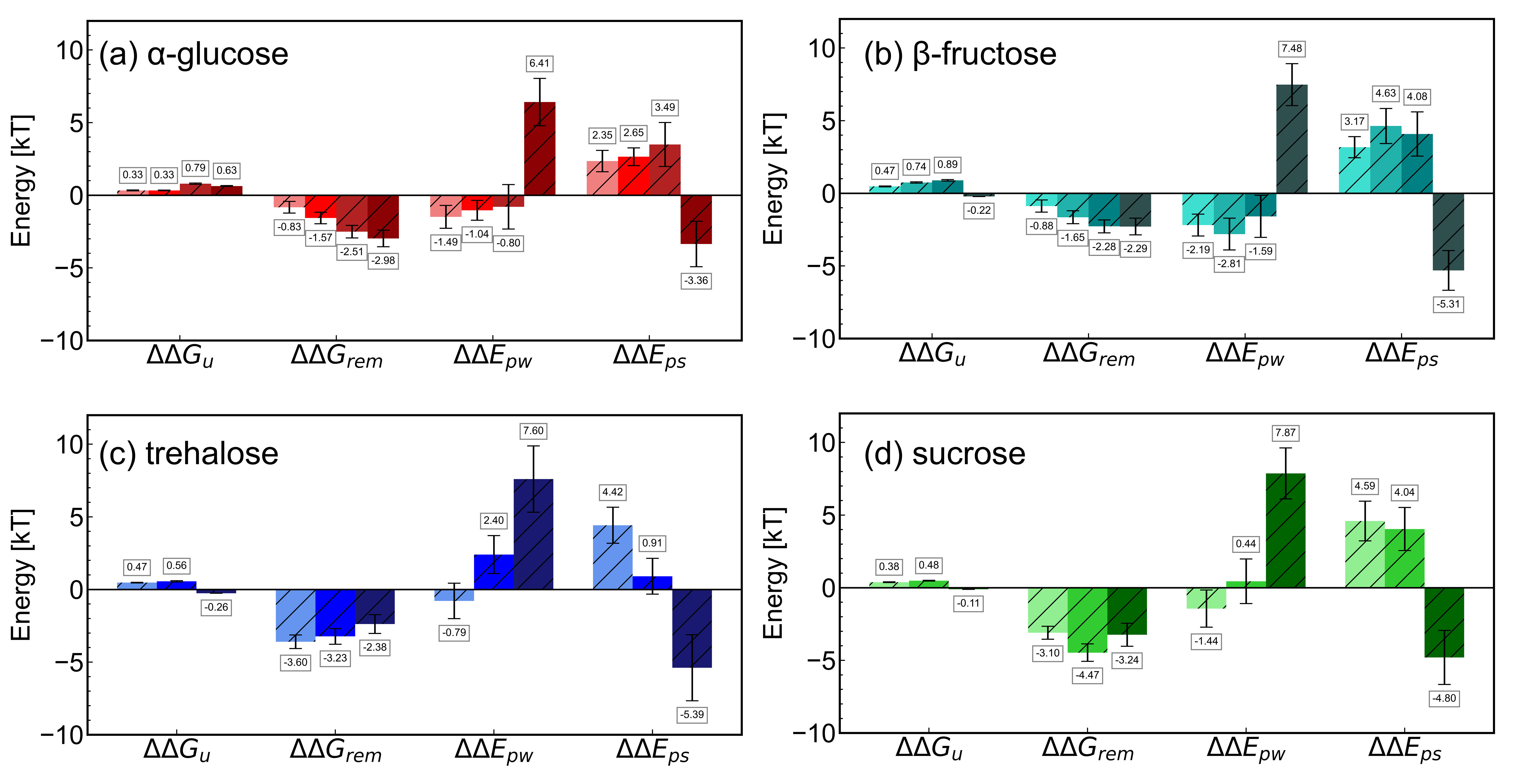}
 \caption{Decomposition of CP unfolding free energy contributions in (a) $\alpha$-glucose, (b) $\beta$-fructose, (c) trehalose, and (d) sucrose solutions. Positive values make the unfolding in the presence of sugar molecules less favorable than in pure water. Increasing additive concentration is denoted by increased shading (light to dark).}
    \label{fig:cp-pmf-decomp}
\end{figure*}

We decompose the polymer unfolding free energy to gain insights into the various contributions underlying sugar effects on polymer stability (as described in Section \ref{sec-decomp}). The contributions to $\Delta\Delta G_u$ ($= \Delta G_{u,sugar-soln} - \Delta G_{u,water}$) are shown in Fig. \ref{fig:hp-pmf-decomp} for HP and Fig. \ref{fig:cp-pmf-decomp} for CP. 

For HP in monosaccharide solutions (Fig. \ref{fig:hp-pmf-decomp}a, b), polymer-sugar interactions ($\Delta \Delta E_{ps}$) stabilize the folded state up to a concentration of 0.5 M. At 1 M, this effect reduces, and by 2 M, these interactions favor the unfolded state. This transition suggests that at higher sugar concentrations, additional binding sites on the polymer that become accessible upon unfolding interact preferentially with sugars, thereby stabilizing the unfolded conformation. This is also largely supported by the increased $\Gamma$ values for the unfolded state at high sugar concentrations. These findings explain the decrease in $\Delta \Delta G_{u}$ observed at high sugar concentrations. Conversely, polymer-water interactions ($\Delta \Delta E_{pw}$) favor unfolding at concentrations up to 0.5 M but promote folding at higher concentrations. $\Delta \Delta G_{rem}$, representing the entropic effects, consistently favors unfolding in all sugar solutions relative to pure water. 

In disaccharide solutions (Fig. \ref{fig:hp-pmf-decomp}c, d), a similar trend is observed, but the stabilizing effect of $\Delta \Delta E_{ps}$ shifts toward favoring the unfolded state as early as 0.5 M. This earlier onset of stabilization relative to monosaccharides can be attributed to the larger molecular size of disaccharides: a 0.5 M disaccharide solution contains approximately the same total sugar mass as a 1 M monosaccharide solution. Consequently, trends that appear at 1 M for monosaccharides tend to emerge at roughly 0.5 M in disaccharide solutions. This mass-equivalent comparison helps explain why $\Delta G_u$ decreases at lower molar concentrations in disaccharides relative to monosaccharides. While this mass-based comparison captures broad mass-dependent trends, deviations among sugars at equal mass concentrations indicate that other molecular features also contribute to the observed behavior, as explained further in Section \ref{sec-int-entropy} and \ref{sec-entropy}.

For CP, trends similar to those observed for HP are observed: polymer-sugar interactions ($\Delta \Delta E_{ps}$) initially stabilize the folded state but shift to favor the unfolded state at higher sugar concentrations. Polymer-water interactions ($\Delta \Delta E_{pw}$) initially favor unfolding and transition to stabilizing the folded state at elevated concentrations, while $\Delta \Delta G_{rem}$ consistently favors the unfolded state across all concentrations. The magnitude of polymer-sugar interactions $\Delta \Delta E_{ps}$ in CP is lower than in HP at equivalent concentrations, in line with lower $\Gamma$ values observed in CP.

As discussed in Section \ref{sec-pmf-decomp}, $\Delta \Delta G_{rem}$ consists of entropic effects from cavity creation (excluded volume contribution) and polymer-solvent interactions (referred to as the interaction entropy contribution). Previous studies suggest that the excluded volume contribution is directly related to solvent surface tension. \cite{timasheff1998control, kita_contribution_1994, arsiccio2023thermodynamic} Specifically, higher surface tension increases the energy required for cavity formation, favoring smaller cavities and thus stabilizing the folded state. Since all sugars considered here increase the air-water surface tension \cite{timasheff1998control, docoslis2000influence, lin1996role, adhikari2007effect}, one would expect $\Delta \Delta G_{rem}$ to promote folding by discouraging large cavity formation. However, our results indicate that $\Delta \Delta G_{rem}$ favors the unfolded state. This suggests that the other entropic effect, polymer-solvent interaction entropy, dominates and counteracts the stabilizing excluded volume contribution. This destabilizing role of interaction entropy is consistent with Olgenblum et al. \cite{olgenblum2023not}, whose findings showed that the polymer-solvent interaction entropy opposes folding for two miniproteins in various sugar solutions, including $\alpha$-glucose, trehalose, and sucrose.

These findings highlight the intricate, concentration-dependent interplay among polymer-sugar and polymer-water interactions, as well as entropic effects in determining polymer stability. While sugars generally promote folding at lower concentrations, the extent of stabilization and the concentration where maximum stabilization is reached depend on the sugar type and the macromolecule. Notably, the presence of partial charges in CP alters the balance of these contributions at different concentrations, highlighting the importance of a nuanced understanding of sugar-mediated stabilization mechanisms and their variability across different polymers and proteins. Building on these observations, we delve deeper into the differences in the contributions across sugar types and investigate potential applications for modulating polymer stability.

\subsection{Sugar-Specific Stabilization Mechanisms: The Role of Interactions and Entropy} \label{sec-int-entropy}

In this section, we examine the differences that arise from sugar structure and connectivity on HP folding, focusing on their molecular interactions and thermodynamic contributions. We first examine the differences in monosaccharides, focusing on $\alpha$-glucose, $\beta$-fructose, and their equimolar mixture ($\alpha$-gluc+$\beta$-fruc). We then compare the thermodynamic contributions of monosaccharides and their corresponding disaccharides.

\begin{figure}[!ht]
 \centering
 \includegraphics[width=1\textwidth]{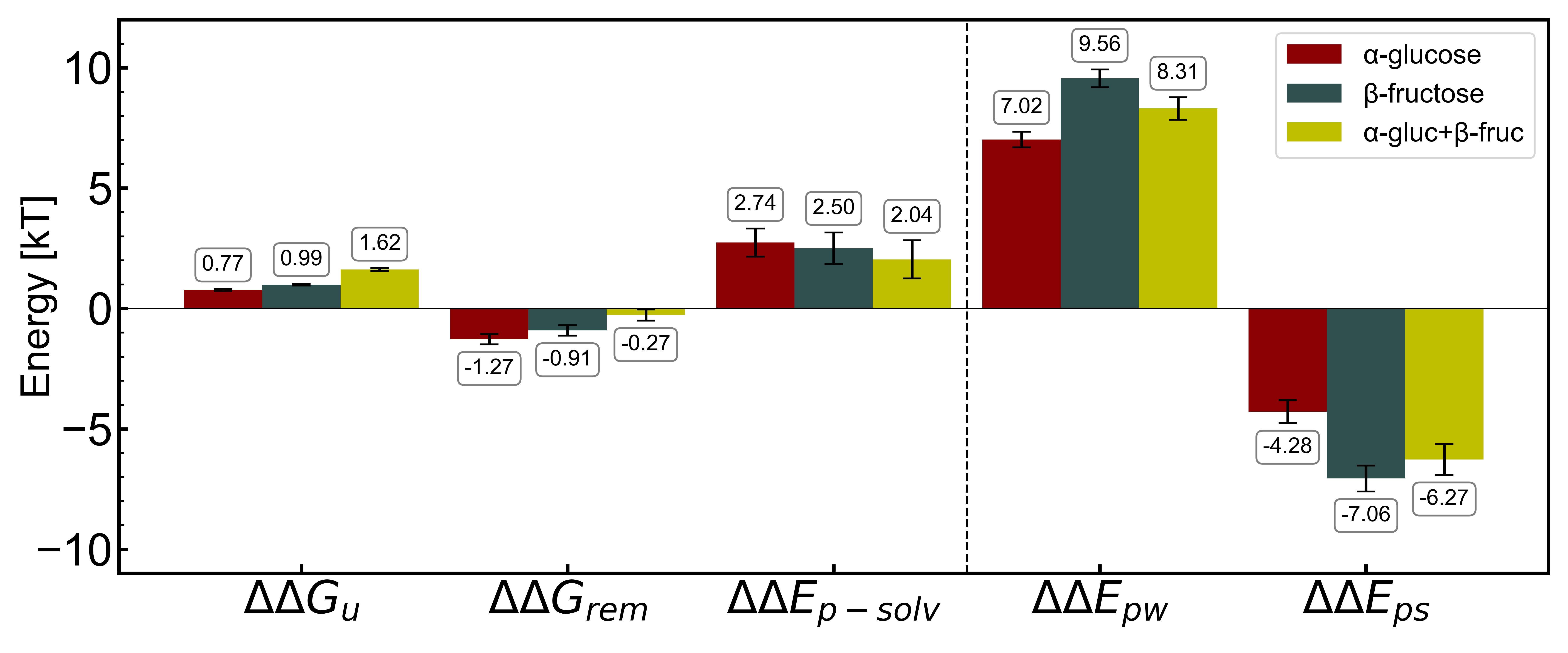}
 \caption{Decomposition of HP unfolding free energy contributions in 2M monosaccharide ($\alpha$-glucose, $\beta$-fructose, $\alpha$-gluc+$\beta$-fruc) solutions. Positive
values favor HP folding and negative values favor HP unfolding.}
    \label{fig:hp-mono}
\end{figure}

The unfolding free energies, shown in Fig. \ref{fig:hp-pmf}e, indicate that $\beta$-fructose promotes folding of HP more effectively than $\alpha$-glucose at equivalent concentrations (0.99 vs 0.74 kT at 0.25 M, 1.41 vs 1.24 kT at 0.5 M, 2.66 vs 1.81 kT at 1 M, and 0.66 vs 0.44 kT at 2 M). To understand the origin of this difference, we compare the thermodynamic decomposition of the PMF obtained for the two sugars. The results of this decomposition for all concentrations are shown in SI Fig. S5. We group the polymer-sugar interactions and polymer-water interactions into a single polymer-solvent interaction term ($\Delta\Delta E_{p-solv}$) as per Eq. \ref{eqn:decompE}. At low concentrations (up to 1 M), $\Delta\Delta E_{p-solv}$ more strongly favor folding in $\beta$-fructose than in $\alpha$-glucose solution (SI Fig. S5). However, at 2 M, $\Delta\Delta E_{p-solv}$ favored folding more in $\alpha$-glucose than in $\beta$-fructose, although the difference was not statistically significant. Despite this, $\beta$-fructose exhibited a higher overall $\Delta \Delta G_u$, indicating greater stabilization of the folded state (Fig. \ref{fig:hp-mono}). This difference is reconciled by considering the $\Delta \Delta G_{rem}$ term, which favored folding more in $\beta$-fructose than in $\alpha$-glucose. Together, these findings illustrate a concentration-dependent stabilization mechanism, wherein polymer-solvent interaction energies differentiate stabilizing excipients at low concentrations, while entropic effects become increasingly important at higher concentrations.

To elucidate the molecular basis of the low-concentration behavior, we decompose the polymer-sugar interaction energies by categorizing the sugars into three mutually exclusive groups: hydroxymethyl ($\mathrm{CH_2OH}$), hydroxyl ($\mathrm{OH}$), and ring backbone groups. This categorizing approach is generalizable across the sugars studied, collectively accounting for all atoms in the sugars, and provides a breakdown of the specific interactions that drive sugar-dependent stabilization. As shown in SI Fig. S6, the additional $\mathrm{CH_2OH}$ group in $\beta$-fructose is the primary contributor to its enhanced stabilization at low concentrations. Contributions from the ring and $\mathrm{OH}$ groups in $\beta$-fructose are comparable to or lower than those in $\alpha$-glucose, consistent with $\beta$-fructose's smaller furanose ring and fewer hydroxyl groups. The subgroup interactions follow a consistent trend across all concentrations: for $\alpha$-glucose, $\mathrm{CH_2OH} < \mathrm{OH} < \text{Ring}$, whereas for $\beta$-fructose, $\mathrm{OH} < \text{Ring} < \mathrm{CH_2OH}$. Radial distribution function analysis (SI Fig. S7) indicates no significant spatial differences in the organization of the $\beta$-fructose's two $\mathrm{CH_2OH}$ around the HP. 

To further probe the high-concentration behavior and the role of entropic contributions, we studied HP in a 2 M equimolar $\alpha$-gluc+$\beta$-fruc mixture. Interestingly, this mixture provided greater stabilization than either pure $\alpha$-glucose or $\beta$-fructose solutions, as shown in Fig. \ref{fig:hp-mono}. Comparing the individual thermodynamic contributions with the pure solutions, the polymer-sugar and polymer-water interaction energies in the $\alpha$-gluc+$\beta$-fruc mixture are bounded by those of the pure solutions. However, the key difference is in the $\Delta \Delta G_{rem}$ term, which favored unfolding less in the mixture compared to the pure solution. This suggests that the mixture enhances the configurational diversity of the local solvent environment, thereby increasing polymer-solvent entropic contributions that favor the folded state.

\begin{figure}[!ht]
 \centering
 \includegraphics[width=1\textwidth]{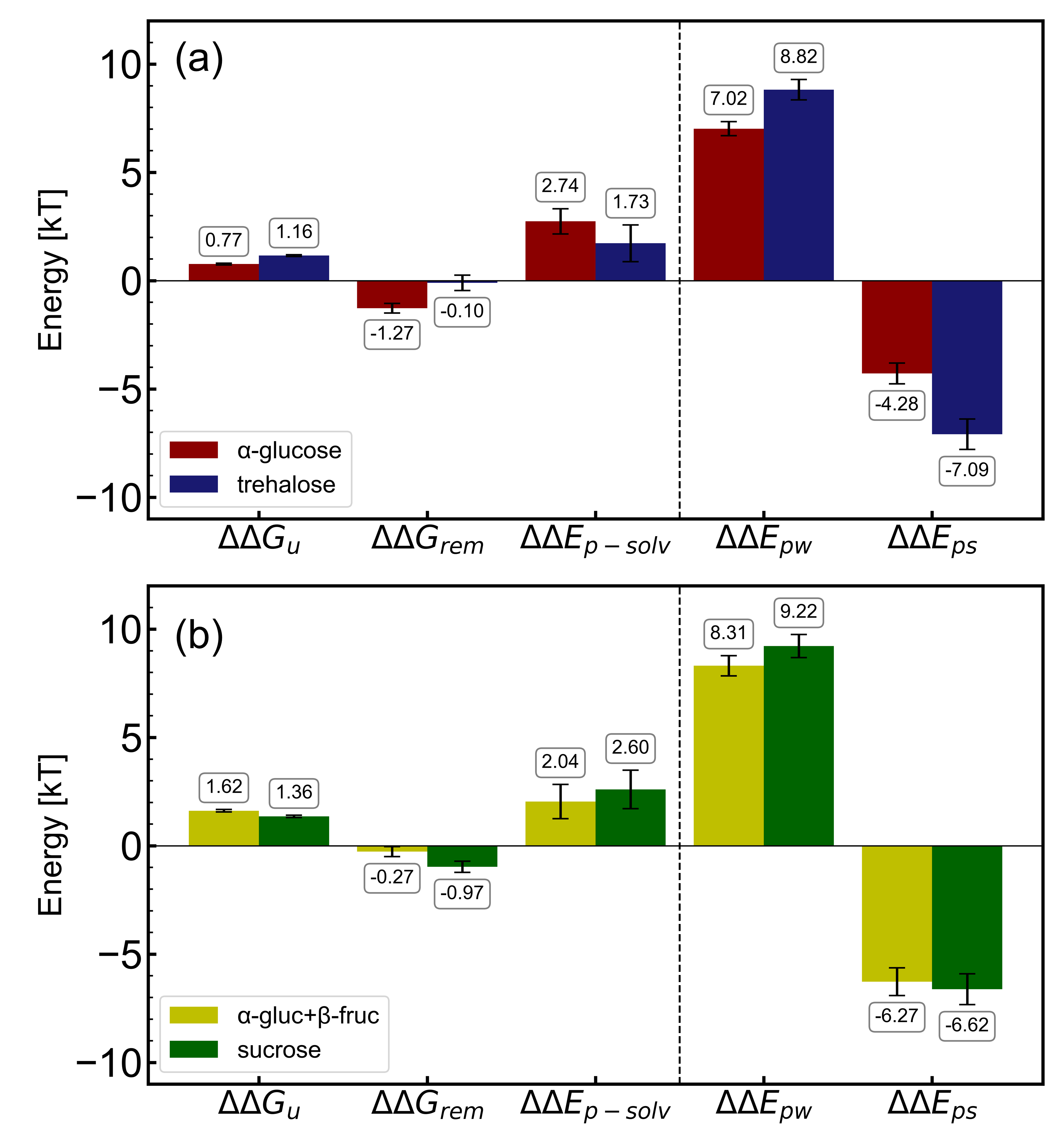}
 \caption{Decomposition of HP unfolding free energy contributions in (a) 2 M $\alpha$-glucose vs 1 M trehalose and (b) 2 M $\alpha$-gluc+$\beta$-fruc vs 1 M sucrose solutions. Positive
values favor HP folding and negative values favor HP unfolding.}
    \label{fig:hp-exc-pmf-decomp-mono-di}
\end{figure}

We next compare HP stabilization and their thermodynamic contributions in monosaccharide and disaccharide solutions (Fig. \ref{fig:hp-exc-pmf-decomp-mono-di}) at equivalent monomer concentrations. At low concentrations (0.5 M monomer concentration) trehalose, a disaccharide of two $\alpha$-glucose units, stabilizes HP more effectively than $\alpha$-glucose, primarily due to polymer-solvent interactions (refer to SI Fig. S8). However, at high concentrations (2 M monomer concentration), trehalose stabilizes HP primarily due to differences in $\Delta \Delta G_{rem}$. Similarly, at high concentrations, 1 M sucrose, composed of $\alpha$-glucose and $\beta$-fructose units, is outperformed by the 2 M $\alpha$-gluc+$\beta$-fruc mixture, again due to changes in $\Delta \Delta G_{rem}$. These observations further highlight the role of polymer-solvent interaction entropy, especially in the high-concentration regime (2 M). In the following section, we compute the local mixing entropy from simulations to evaluate its utility as a proxy for polymer-solvent interaction entropy.

\subsection{Local Mixing Entropy as a Proxy for Entropic Contributions} \label{sec-entropy}

To assess the role of entropy in polymer stabilization at high sugar concentrations, we compute a simulation-based metric inspired by the solution mixing entropy. Specifically, we estimate the local mixing entropy of solvent components surrounding the polymer as a proxy for polymer-solvent interaction entropy.

This metric is evaluated by defining a local solvent shell around the polymer, consistent with the region used to compute preferential interaction coefficients. Within this shell, we calculate the atom number fractions of water and sugar atoms in each simulation frame, ensuring that each atom is uniquely identified and counted only once, even in geometrically overlapping shell regions. Because these fractions are normalized, $\Delta\Delta S_{mix}$ is independent of the total shell volume or the number of local atoms, reflecting only changes in the local compositional diversity upon folding. Based on these fractions, we compute the local mixing entropy as described in Section~\ref{section-local-mixing-entropy}, and determine the entropy difference between the folded and unfolded states, denoted as $\Delta\Delta S_{mix} = \Delta S_{mix, unfolded} - \Delta S_{mix, folded}$. Positive values of $\Delta\Delta S_{mix}$ indicate greater solvent compositional heterogeneity—and thus higher configurational entropy—in the unfolded state.

In our thermodynamic decomposition, $\Delta S_{ex}$ represents the entropy change associated with depletion due to excluded volume, and $\Delta S_{si}$ corresponds to configuration-specific interaction effects. In contrast, $\Delta\Delta S_{mix}$ captures compositional changes in the local solvent environment, which may arise from either excluded-volume–driven depletion or specific sugar–polymer interactions. Therefore, this metric quantifies the compositional entropy component of polymer stabilization, rather providing a measure of the number of local atoms or the total configurational entropy.

\begin{figure}[!ht]
 \centering
 \includegraphics[width=0.75\textwidth]{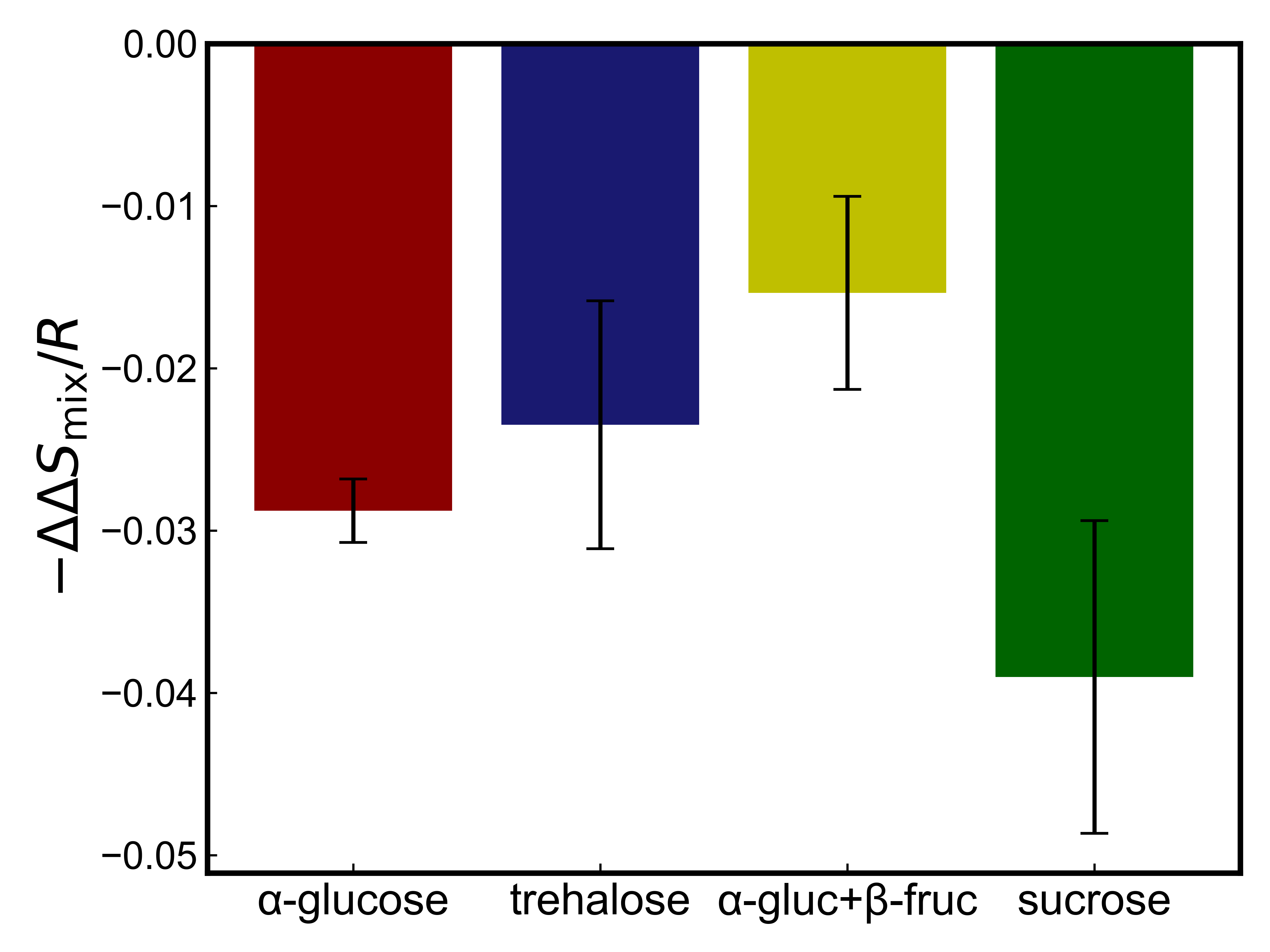}
 \caption{Difference in local mixing entropy (-$\Delta\Delta S_{\text{mix}}$) between folded and unfolded states of HP in high-concentration sugar solutions (2 M monosaccharide or 1 M disaccharide solutions). Positive values in the plot indicate higher mixing entropy in the folded state, suggesting enhanced configurational diversity of the local solvent environment in the folded state compared to the unfolded state.}
    \label{fig:delta-entropy-highconc-localrad7}
\end{figure}

We compare the $\Delta \Delta S_{mix}$ across high-concentration monosaccharide and their corresponding disaccharide solutions, to isolate the contribution of local solvent mixing entropy and evaluate its utility as a proxy for the polymer-solvent interaction entropy. Fig. \ref{fig:delta-entropy-highconc-localrad7} shows the $\Delta \Delta S_{mix}$ calculated from a local radius of 7 \AA. The results observed are unchanged with variations in the local radius from 5 \AA ~to 10 \AA ~as shown in SI Fig. S9.  From Fig. \ref{fig:delta-entropy-highconc-localrad7}, we see that 1 M trehalose has a lower $\Delta \Delta S_{mix}$ than 2 M $\alpha$-glucose and $\Delta \Delta S_{mix}$ of 2 M $\alpha$-gluc+$\beta$-fruc is lower than that of 1 M sucrose. Larger (more positive) $\Delta\Delta S_{mix}$ values correspond to a greater tendency for unfolding, indicating that sucrose promotes unfolding the most, while the $\alpha$-glucose + $\beta$-fructose mixture promotes it the least. These results complement our thermodynamic decomposition by differentiating the extent of entropic contributions from various excipients at high concentrations. 

We further analyzed $S_{mix}$ as a function of the polymer radius of gyration ($R_g$) to examine the local compositional entropy around the stable polymer conformations. This analysis revealed that the hairpin configuration (state II) exhibits $S_{mix}$ values comparable to or slightly higher than those of the fully collapsed state (state I). At low concentrations (sugar atom number fraction $x_s$ < 0.5), this indicates similar or greater local sugar enrichment in the hairpin state. At higher concentrations ($x_s$ > 0.5), the $x_s$ (state I) > $x_s$ (state II) resulting in higher $S_{mix}$ in State II. This happens because at higher concentrations, with the increase in $R_g$, while the number of sugar atoms in the vicinity of the polymer increases, the increase in water atoms is higher. We hypothesize that this could be related to the smaller size of water molecules being able to penetrate the already crowded solvent shell resulting in an increase in mixing entropy. Representative $S_{mix}$–$R_g$, $x_s$-$R_g$, $n_s$-$R_g$, and $n_w$-$R_g$ plots illustrating these trends are provided in the SI Fig. S10.

% Thus, we infer that $\Delta \Delta S_{mix}$ captures essential features of polymer–solvent interaction entropy and offers a physically interpretable and computationally efficient measure to assess the entropic component of polymer stabilization.

Overall, $\Delta\Delta S_{mix}$ captures essential features of polymer-solvent interaction entropy and offers a physically interpretable and computationally efficient measure to assess the entropic component of polymer stabilization. These observations build upon previous studies, which have primarily attributed protein stability to excluded volume effects and specific protein-sugar interactions \cite{timasheff1998control, arsiccio2023thermodynamic, olgenblum2023not}. Our findings highlight polymer-solvent interaction entropy as an additional, previously underexplored strategy for enhancing stability in biological formulations. The interplay of direct interactions, excluded volume, and local configurational entropy collectively governs the stabilizing capacity of sugar solutions, with different contributions being decisive at different concentrations. Future studies that systematically manipulate polymer-solvent interaction entropy - through sugar mixtures, excipient design, or solvent composition - will be key to determining its broader applicability in optimizing sugar-based excipients for biological formulations.

\section*{Conclusions}
This study investigates sugar-mediated stabilization of HP and CP models, which serve as simplified systems for understanding protein behavior. Analysis of the PMF reveals concentration-dependent stabilization, where increasing sugar concentration initially enhances polymer stability, followed by a relative decrease in stabilization at higher concentrations. The concentration at which this change occurs is lower for disaccharides than for monosaccharides, suggesting the role of molecular size in impacting this behavior.

Preferential interaction coefficient analysis shows that sugars preferentially interact with both folded and unfolded polymer states, with disaccharides exhibiting stronger interactions than monosaccharides. Rather than an absolute threshold of $\Gamma$, it is the difference in preferential interactions between the two states ($\Delta \Gamma$) and how this evolves with concentration that determines the change in unfolding free energy. Free energy decomposition further reveals the distinct roles of polymer-sugar interaction energy, polymer-water interaction energy, and entropic contributions to polymer stabilization. Polymer-sugar interactions favor stabilization at low sugar concentrations but transition to favoring unfolding at higher concentrations. Polymer-water interactions, on the other hand, favor unfolding at low sugar concentrations but transition to favoring folding at higher concentrations. The entropic effects consistently favor unfolding across all sugar solutions. Since sugars considered in this study increase the surface tension of water, we expect excluded volume contribution in the entropic effects to favor folding and propose that the polymer-solvent interaction entropy is the contribution that favors unfolding.

By comparing monosaccharides and disaccharides, we identify that sugar-specific effects arise from changes in both direct interactions and polymer-solvent entropic effects. Changes in polymer-solvent interactions differentiate stabilizing effects at low concentrations, while changes in polymer-solvent interaction entropy differentiate stabilizing effects at high concentrations. Notably, we observe that the $\alpha$-gluc+$\beta$-fruc mixture can exploit this interaction entropy such that it stabilizes the folded polymer better than the corresponding pure monosaccharide or disaccharide solutions. We further demonstrate that local solvent mixing entropy serves as a proxy for polymer-solvent interaction entropy and could differentiate between stabilizing and destabilizing conditions, especially when interaction energies are comparable.

Overall, our results underscore the complexity of protein stabilization mechanisms in sugar solutions and emphasize the importance of considering concentration-dependent effects, the nature of sugar-protein interactions, and the protein-solvent entropy when designing formulations. These findings contribute to a more nuanced understanding of excipient-protein interactions, which can inform more effective formulation strategies in biopharmaceutical design. Additionally, the framework presented here is also relevant to understanding crowding effects on protein folding equilibria in cellular environments. While the present work focused on the (++––) charge sequence, exploring alternative arrangements could further elucidate the influence of charge distribution on polymer folding and excipient-mediated stabilization. Future work will also aim to extend these findings to protein systems with diverse surface chemistries and folding behaviors to develop a more generalized understanding of sugar-mediated stabilization mechanisms.

\section{Supporting Information}

Supporting information includes KBP force field validation results, additional simulation details, convergence check for PMFs, variation of preferential interaction coefficient with local domain cutoff radius, unfolding free energy vs  change in preferential interactions, PMF decomposition and subgroup interaction energies, radial distribution function of $\beta$-fructose $\mathrm{CH_2OH}$ groups around the HP, change in mixing entropy with local radius, PMF decomposition results with replicate runs. 

\begin{acknowledgement}
This material is based upon work supported by the National Science Foundation under DMREF Grant No. 2325392, 2118788, 2118693, and 2118638. The Minnesota Supercomputing Institute at the University of Minnesota Twin Cities provided computational resources.
\end{acknowledgement}

\bibliography{refs}

\end{document}

\setcounter{secnumdepth}{4}

\renewcommand{\thefigure}{S\arabic{figure}}
\setcounter{figure}{0}

\renewcommand{\thetable}{S\arabic{table}}
\setcounter{table}{0}

\renewcommand{\theequation}{S\arabic{equation}}
\setcounter{equation}{0}

\renewcommand{\thepage}{S\arabic{page}}

\emergencystretch 3em

\section{KBP Force Field Validation}

We use the Kirkwood-Buff (KB) integral-based carbohydrate parameters developed by Cloutier et al. \cite{cloutier2018kirkwood} to parameterize monosaccharides and disaccharide molecules and the TIP4P/2005 water model. However, the original KB parameters were developed to be compatible with the TIP3P water model\cite{jorgensen1983comparison}. To analyze the extent of this change in the water model, we computed the KB integral values of water-water, sugar-sugar, and sugar-water for 0.5m and 1m trehalose and validated against the values obtained by Cloutier et al. and experimental data (shown in table \ref{tbl:si-kbp}). Unlike the main text, which reports concentrations in molarity, we use molality here to ensure consistency with Cloutier et al. \cite{cloutier2018kirkwood}. Simulations were performed using the same box length and number of molecules as in their study. Overall, we observed small deviations from the KB values computed by Cloutier et al.\cite{cloutier2018kirkwood} but still within the range of experimental error.

\begin{table*}
  \caption{Comparison of KB Integral values}
  \label{tbl:si-kbp}
  \begin{tabular}{c c c c c}
    \hline
    System & Model & $G_{11}$ & $G_{33}$ & $G_{31}$ \\
    \hline
     & KBP + TIP3P & 1 & -845 $\pm$ 50 & -180 $\pm$ 5 \\
    0.5m trehalose & KBP+TIP4P/2005 & 2.036 $\pm$ 0.33 & -870 $\pm$ 0.31 & -162 $\pm$ 0.06 \\
     & Expt & 4.5 $\pm$ 10 & -625 $\pm$ 220 & -199 $\pm$ 35 \\
     \hline
    & KBP + TIP3P & 14.5 & -670 $\pm$ 30 & -156 $\pm$ 4 \\
    1m trehalose & KBP+TIP4P/2005 & 16 $\pm$ 0.7 & -590 $\pm$ 0.13 & -152 $\pm$ 0.1 \\
     & Expt & 21 $\pm$ 19 & -560 $\pm$ 120 & -169 $\pm$ 37 \\
    \hline
  \end{tabular}
\end{table*}

\section{Simulation Details}

A cubic box of length 6.74 $\mathrm{nm}$ was constructed with a padding of 1.5 $\mathrm{nm}$ between the edge of the fully extended polymer and the nearest box edge. Lorentz-Berthelot mixing rules\cite{lorentz_1881, berthelot1898melange} were used to calculate non-bonded interactions between different atom types, except polymer-water oxygen interactions.\cite{zajac2025flipping} 

\begin{table*}
  \caption{Setup of simulated systems. Simulation times for REUS simulations are reported as $W \times S$, which represent the number of windows per replica (W), and the simulation length in each window (S), respectively.}
  \label{tbl:si-simsetup}
  \begin{tabular}{c c c c c}
    \hline
    System & Simulation Time (ns) & Concentration (M) & $N_{Exc}$ & $N_{Wat}$ \\
    \hline
    HP+ TIP4P/05 water & 12 x 100 & 0.00 & 0 & 10188 \\
    HP + $\alpha$-glucose & 12 x 100 & 0.25 & 47 & 9751 \\
    HP + $\alpha$-glucose & 12 x 100 & 0.50 & 93 & 9300 \\
    HP + $\alpha$-glucose & 12 x 100 & 1.0 & 185 & 8369 \\
    HP + $\alpha$-glucose & 12 x 100 & 2.0 & 370 & 6061 \\
    HP + $\beta$-fructose & 12 x 100 & 0.25 & 47 & 9757 \\
    HP + $\beta$-fructose & 12 x 100 & 0.50 & 93 & 9287 \\
    HP + $\beta$-fructose & 12 x 100 & 1.0 & 185 & 8329 \\
    HP + $\beta$-fructose & 12 x 100 & 2.0 & 370 & 5992 \\
    HP + trehalose & 12 x 100 & 0.25 & 47 & 9403 \\
    HP + trehalose & 12 x 100 & 0.50 & 93 & 8328 \\
    HP + trehalose & 12 x 100 & 1.0 & 185 & 5798 \\
    HP + sucrose & 12 x 100 & 0.25 & 47 & 9402 \\
    HP + sucrose & 12 x 100 & 0.50 & 93 & 8498 \\
    HP + sucrose & 12 x 100 & 1.0 & 185 & 6225 \\
    HP + $\alpha$-glucose + $\beta$-fructose & 12 x 100 & 2.0 & 185+185 & 6077 \\
    \hline
    CP+ TIP4P/05 water & 12 x 100 & 0.00 & 0 & 10188 \\
    CP + $\alpha$-glucose & 12 x 100 & 0.25 & 47 & 9765 \\
    CP + $\alpha$-glucose & 12 x 100 & 0.50 & 93 & 9323 \\
    CP + $\alpha$-glucose & 12 x 100 & 1.0 & 185 & 8364 \\
    CP + $\alpha$-glucose & 12 x 100 & 2.0 & 370 & 6028 \\
    CP + $\beta$-fructose & 12 x 100 & 0.25 & 47 & 9720 \\
    CP + $\beta$-fructose & 12 x 100 & 0.50 & 93 & 9321 \\
    CP + $\beta$-fructose & 12 x 100 & 1.0 & 185 & 8336 \\
    CP + $\beta$-fructose & 12 x 100 & 2.0 & 370 & 5899 \\
    CP + trehalose & 12 x 100 & 0.25 & 47 & 9375 \\
    CP + trehalose & 12 x 100 & 0.50 & 93 & 8449 \\
    CP + trehalose & 12 x 100 & 1.0 & 185 & 5774 \\
    CP + sucrose & 12 x 100 & 0.25 & 47 & 9393 \\
    CP + sucrose & 12 x 100 & 0.50 & 93 & 8434 \\
    CP + sucrose & 12 x 100 & 1.0 & 185 & 6165 \\
    CP + $\alpha$-glucose + $\beta$-fructose & 12 x 100 & 2.0 & 185+185 & 5948 \\
    \hline
  \end{tabular}
\end{table*}

\section{Convergence of PMFs}

\begin{figure*}[!ht]
 \centering
 \includegraphics[width=1\textwidth]{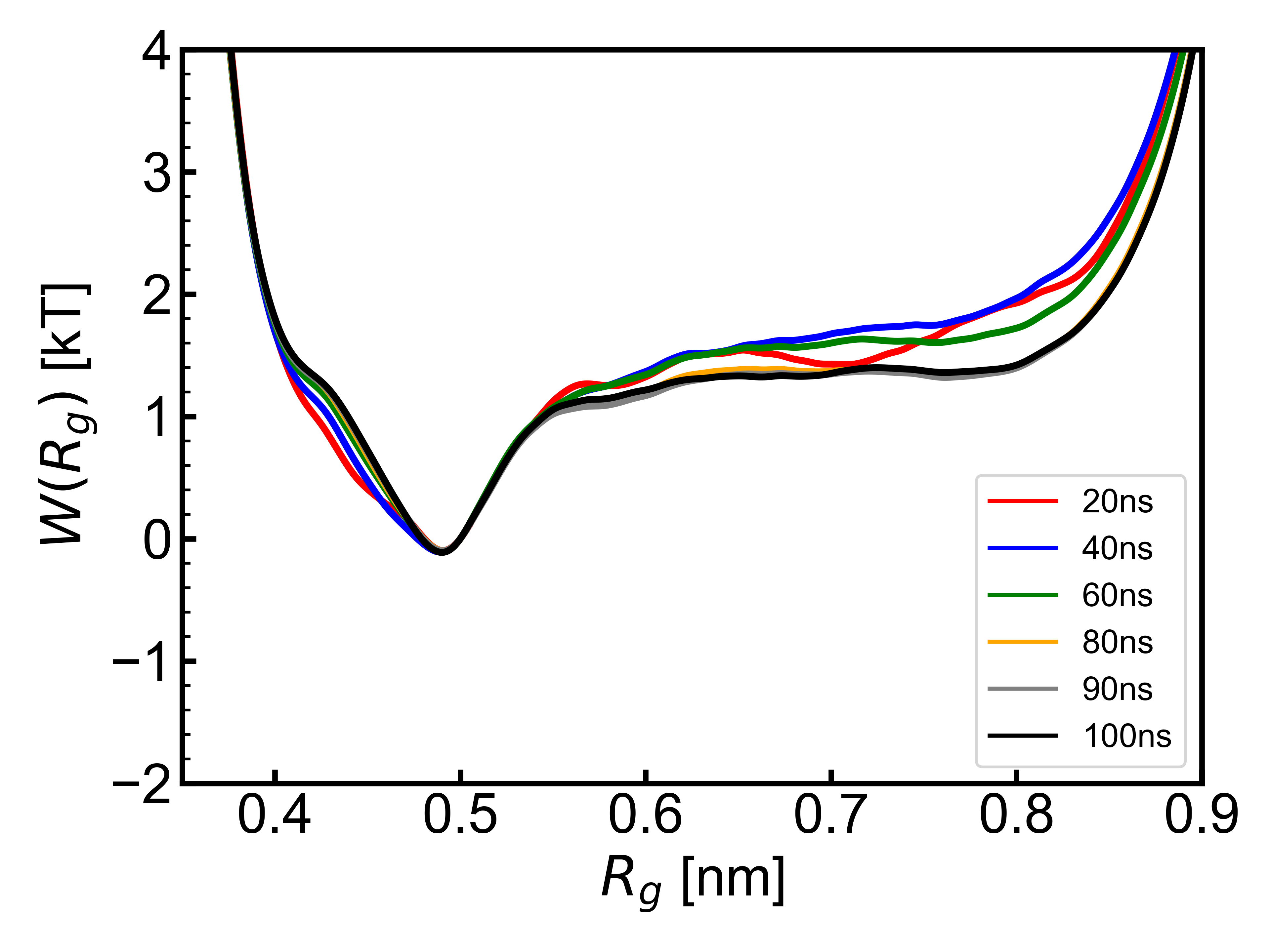}
 \caption{Evolution of PMF with time for the HP in 0.25 M trehalose solution.}
\label{fig:si-pmfevol}
\end{figure*}

To make sure that the PMFs that we computed through simulations have converged, we observed the time evolution of the PMF. Fig. \ref{fig:si-pmfevol} shows the time evolution of PMF of the HP in 0.25 M trehalose solution along the $R_g$ coordinate.

\section{Preferential Interactions vs Local-Bulk Domain Cutoff}

\begin{figure}[H]
 \centering
 \includegraphics[width=1\textwidth]{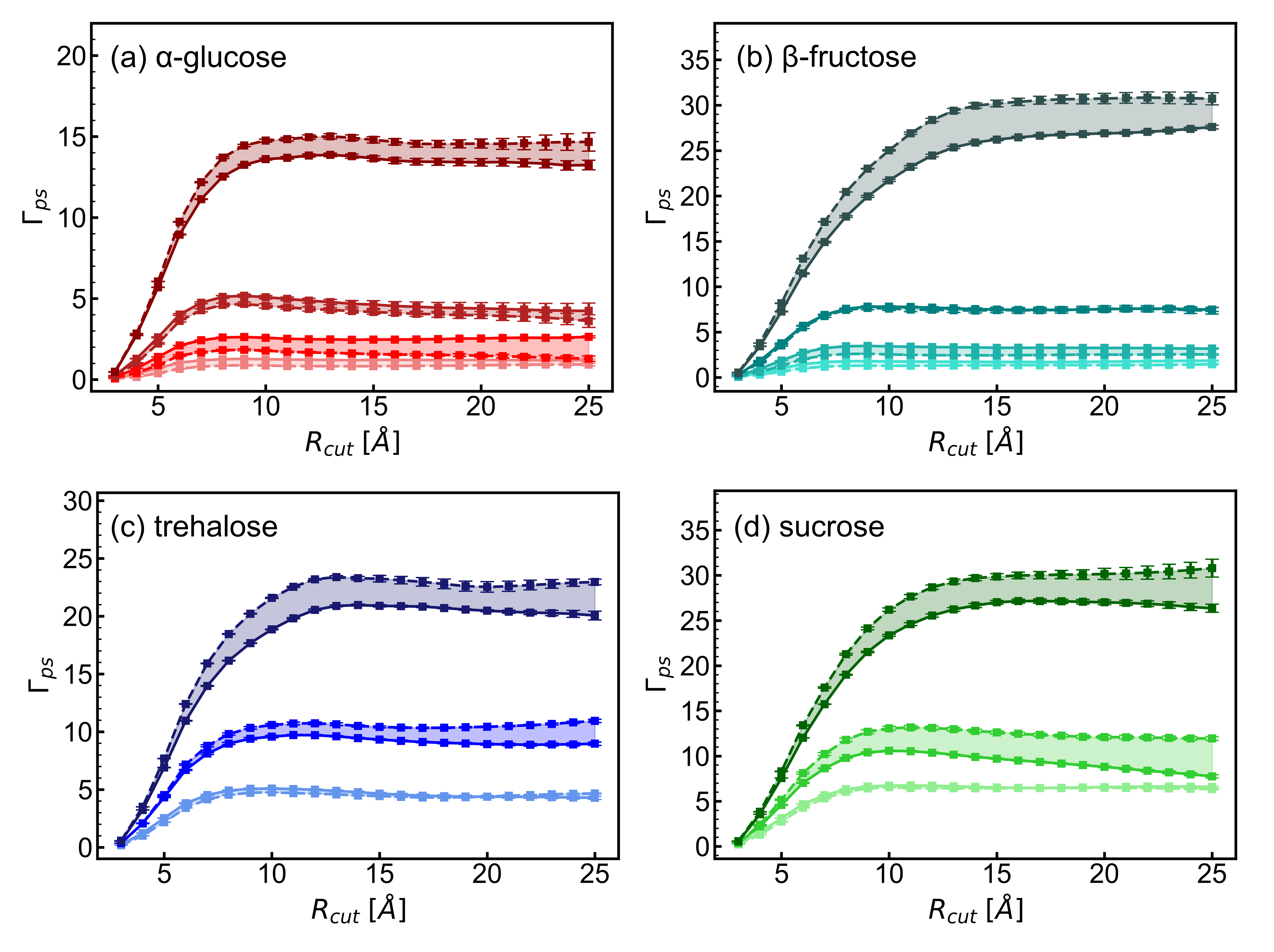}
 \caption{Change in preferential interaction coefficient ($\Gamma_{ps}$ with local-bulk domain cutoff radius ($R_{cut}$) for HP in different sugar solutions.  Red – glucose, blue – trehalose, green – sucrose, cyan - fructose. Increasing shading corresponds to increasing concentration of excipients.}
    \label{fig:si-hp-pref-int}
\end{figure}

\begin{figure}[H]
 \centering
 \includegraphics[width=1\textwidth]{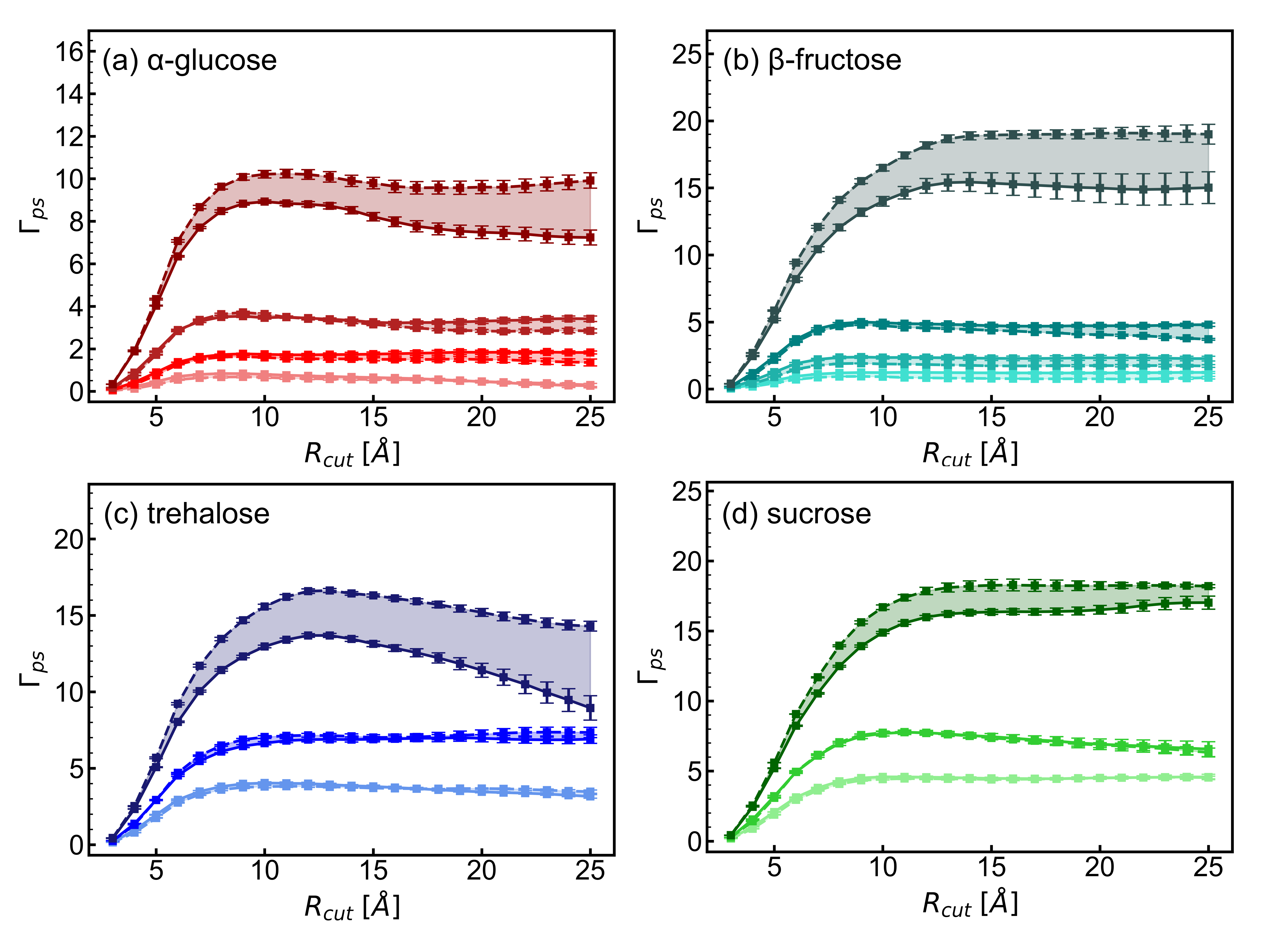}
 \caption{Change in preferential interaction coefficient ($\Gamma_{ps}$ with local-bulk domain cutoff radius ($R_{cut}$) for CP in different sugar solutions.  Red – glucose, blue – trehalose, green – sucrose, cyan - fructose. Increasing shading corresponds to increasing concentration of excipients.}
\label{fig:si-cp-pref-int}
\end{figure}

\section{Free Energy of Unfolding vs Preferential Interactions}

\begin{figure}[H]
 \centering
 \includegraphics[width=1\textwidth]{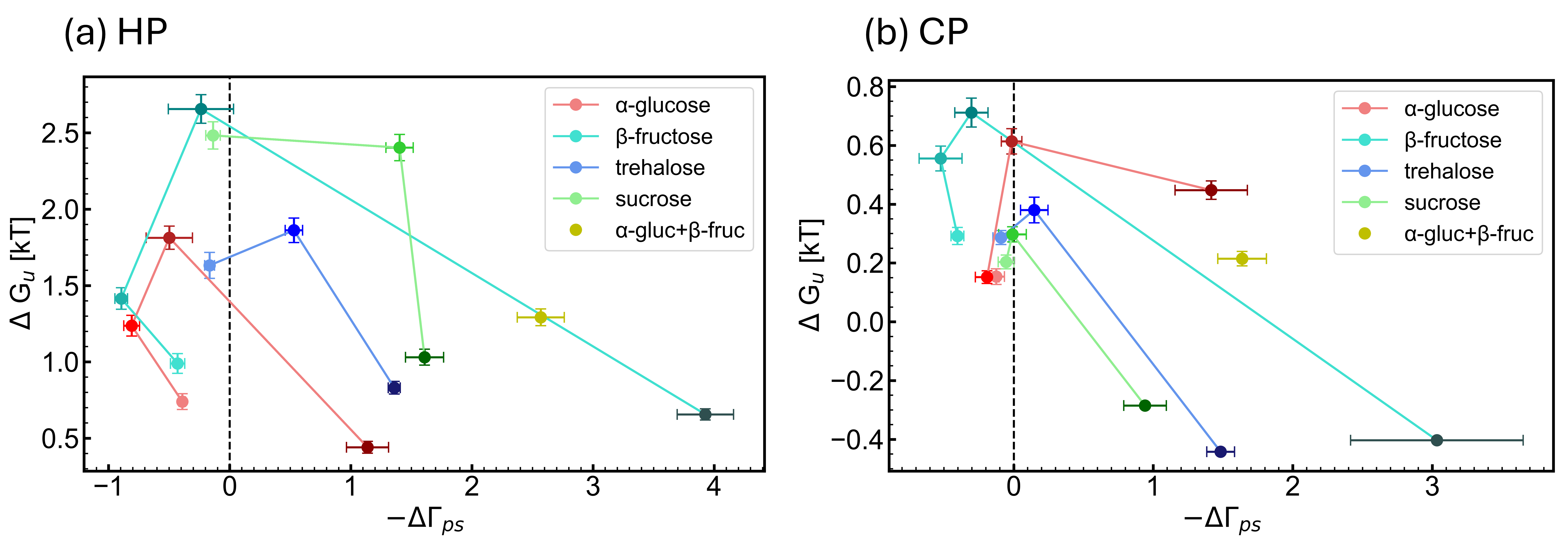}
 \caption{Change in unfolding free energy with change in preferential interaction coefficient ($\Delta \Gamma_{ps}$ for (a) HP and (b) CP in different sugar solutions. Increasing shading corresponds to increasing concentration of excipients.}
\label{fig:si-dg-dpref-int}
\end{figure}

From the figures \ref{fig:si-dg-dpref-int}a and \ref{fig:si-dg-dpref-int}b, we observe that the free energy of unfolding increases when $\Delta \Gamma_{ps}$ is positive and decreases when  $\Delta \Gamma_{ps}$ is negative. In the case of CP in 0.5 M trehalose and 0.5 M sucrose, the preferential interaction coefficients suggest a decreasing trend in the unfolding free energy with increasing sugar concentration. However, this trend reflects the local slope—that is, the direction of change—rather than the absolute value of the free energy. Therefore, it does not necessarily mean that the unfolding free energy at 0.5 M is lower than at 0.25 M. Instead, it means that around 0.5 M, the free energy is decreasing with concentration relative to concentrations just below 0.5 M. Because the jump from 0.25 M to 0.5 M is a finite change rather than an infinitesimal one, we cannot directly infer that the free energy at 0.5 M is lower than at 0.25 M based solely on this trend.

\section{PMF Decomposition: $\alpha$-glucose vs $\beta$-fructose}
%% Take this figure to the SI
\begin{figure}[H]
 \centering
 \includegraphics[width=1\textwidth]{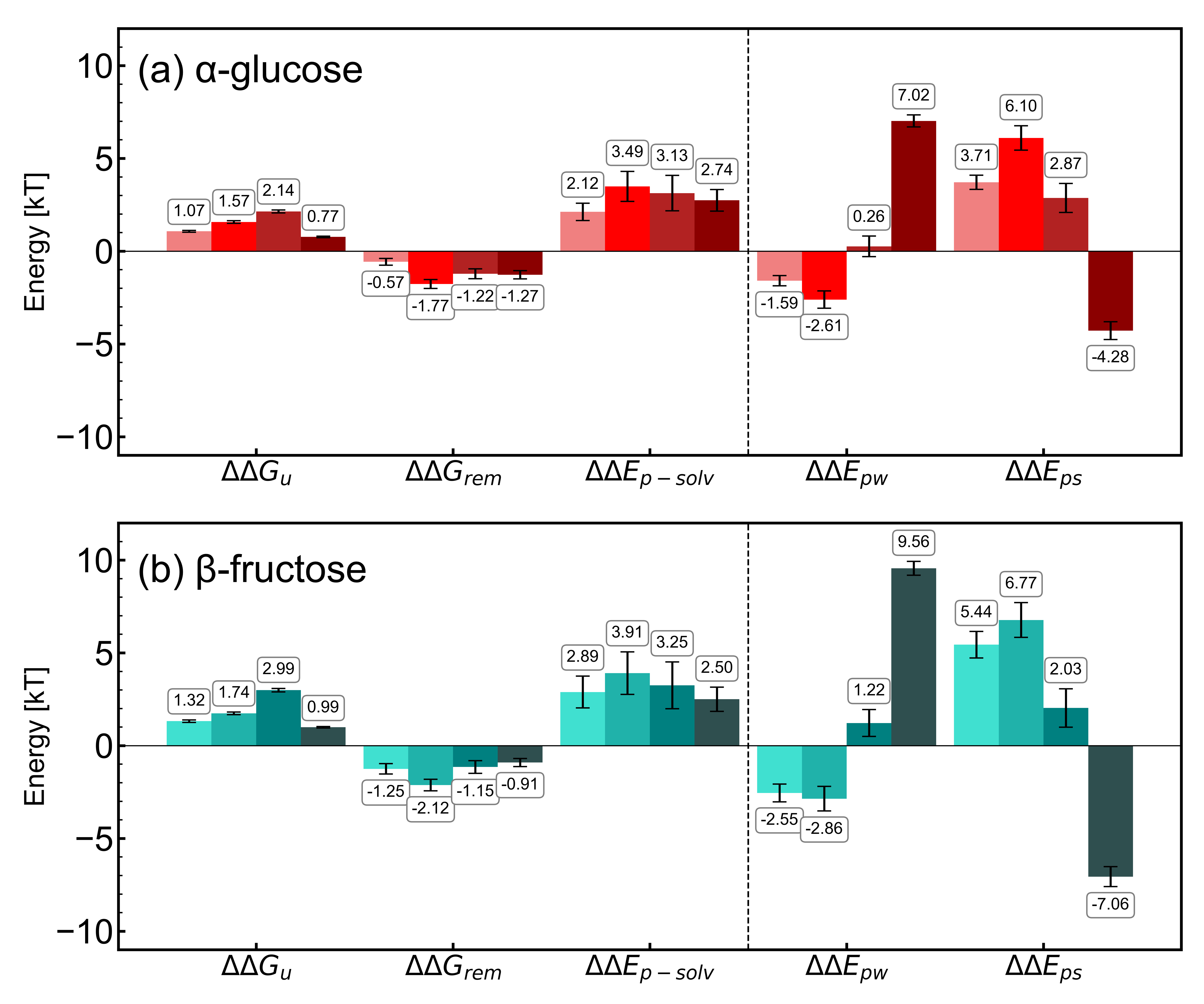}
 \caption{Decomposition of HP unfolding free energy in different concentration solutions of (a) $\alpha$-glucose and (b) $\beta$-fructose. Positive values favor HP folding and negative values favor HP unfolding.}
    \label{fig:si-aglc-bfru-pmf-decomp}
\end{figure}

\section{Subgroup Interaction Energies: $\alpha$-glucose vs $\beta$-fructose}
%% Take this figure to the SI
\begin{figure}[H]
 \centering
 \includegraphics[width=1\textwidth]{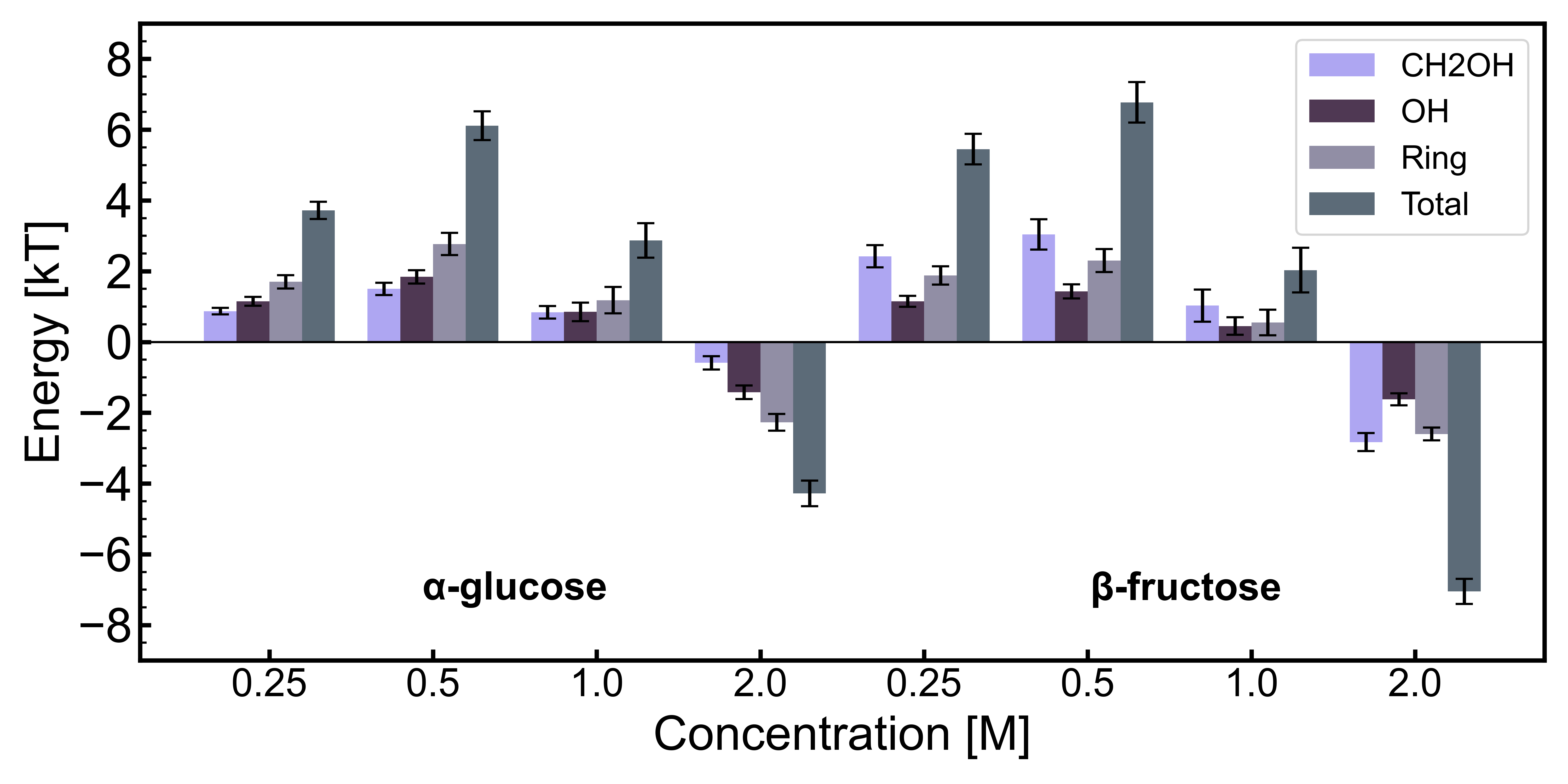}
 \caption{Difference in interaction energies of components with the HP's unfolded and folded states. Positive values mean that the component interacts more with the HP folded state and negative values favors HP unfolding.}
    \label{fig:si-finecomp-aglc-bfru}
\end{figure} 

\section{Radial Distribution Functions of $\beta$-fructose $\mathbf{CH_2OH}$ Groups Around the HP}
\begin{figure}[H]
 \centering
 \includegraphics[width=\textwidth]{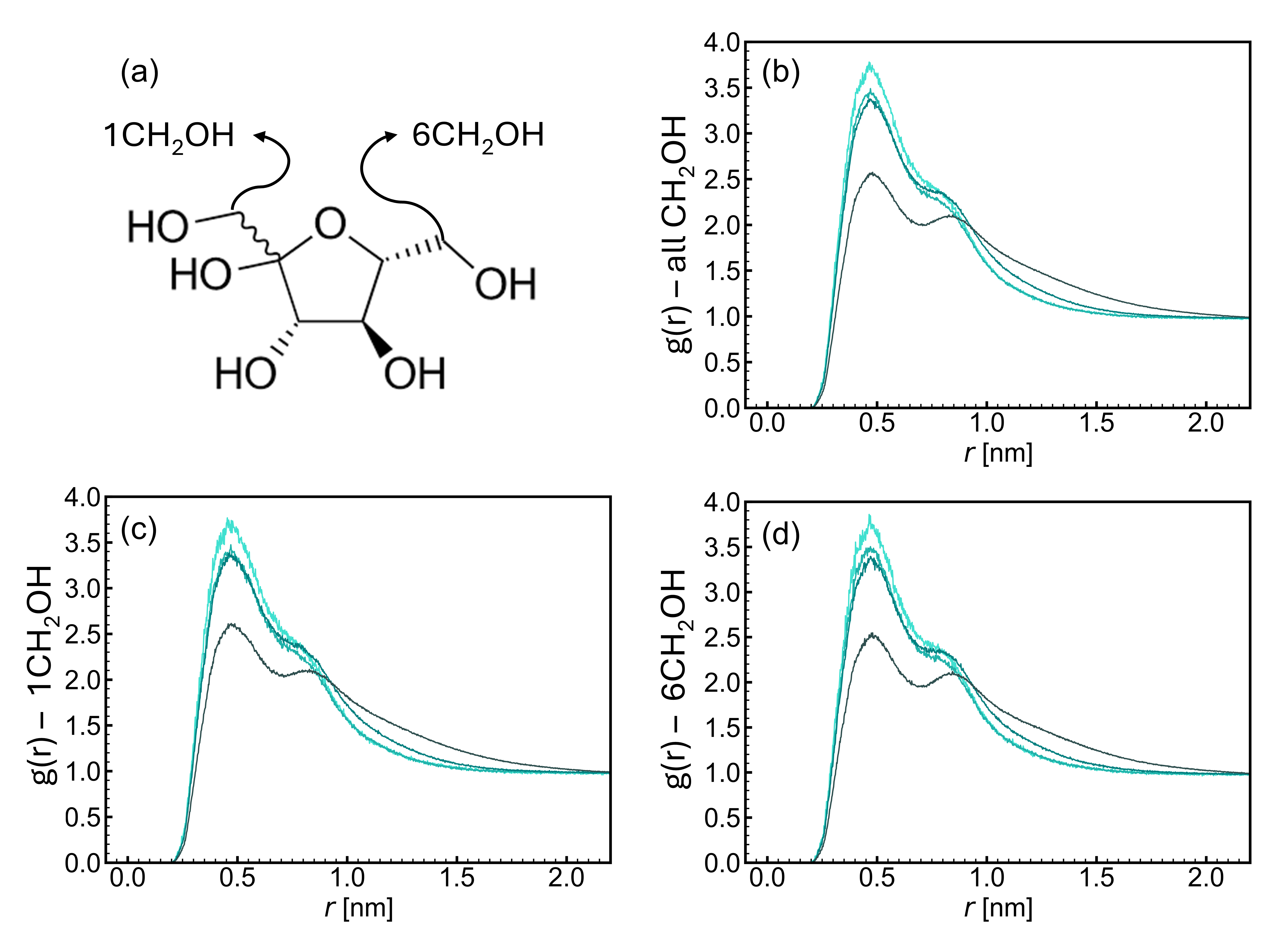}
 \caption{Radial distribution functions of fructose $\mathrm{CH_2OH}$ groups around the HP at different concentrations. Increasing shading corresponds to increasing concentrations. (a) A 2D structure of fructose indicating the two groups: $\mathrm{1CH_2OH}$ and $\mathrm{6CH_2OH}$. RDFs of (b) all $\mathrm{CH_2OH}$, (c) $\mathrm{1CH_2OH}$ and (d) $\mathrm{6CH_2OH}$ around the HP.}
    \label{fig:si-fructose-rdf}
\end{figure}

\section{PMF Decomposition: $\alpha$-glucose vs trehalose at Low and High Concentrations}

\begin{figure}[H]
 \centering
 \includegraphics[width=1\textwidth]{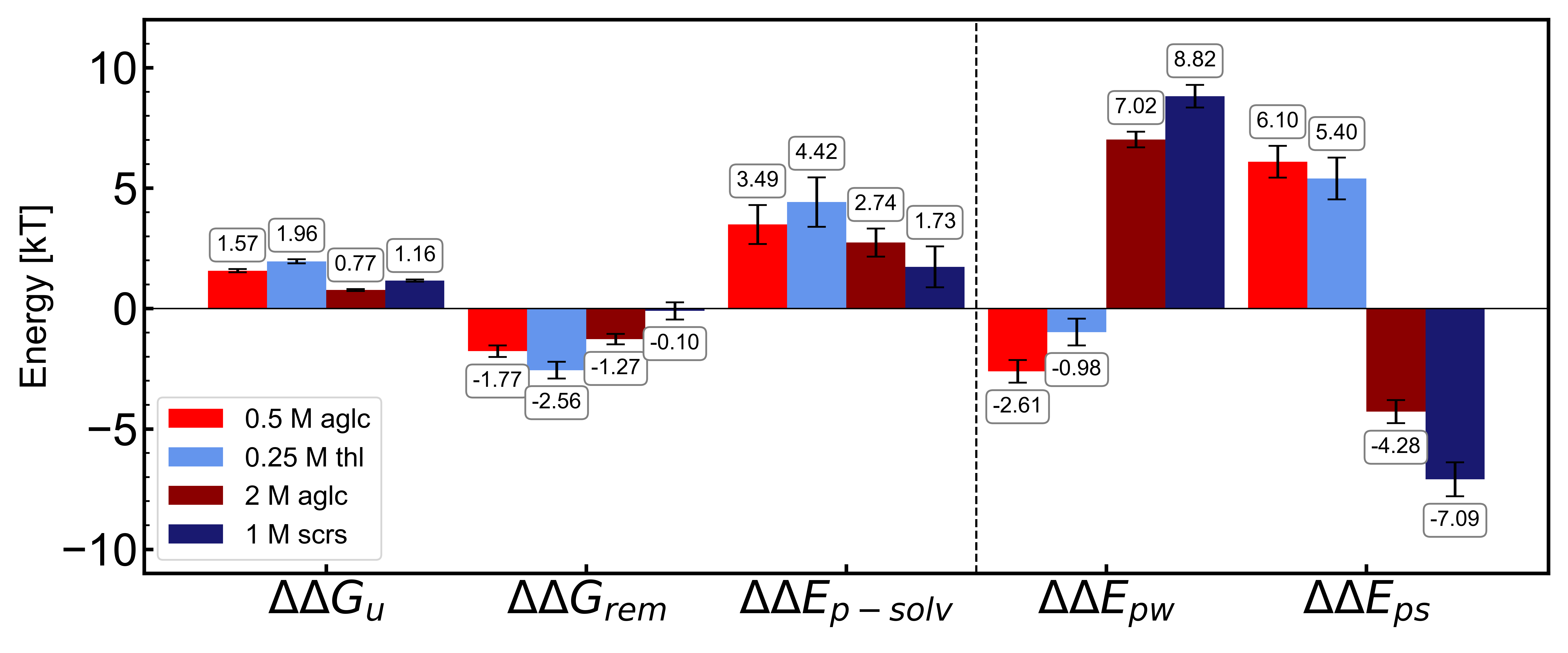}
 \caption{Decomposition of HP unfolding free energy in different concentrations of $\alpha$-glucose and trehalose. Positive values favor HP folding and negative values favor HP unfolding. We clearly see that at low concentrations (0.5 M $\alpha$-glucose vs 0.25 M trehalose), polymer-solvent interaction differences make trehalose a better stabilizer, whereas at high concentrations (2 M $\alpha$-glucose vs 1 M trehalose), $\Delta \Delta G_{rem}$ differences make trehalose a better stabilizer. }
    \label{fig:si-aglc-thl-decomp}
\end{figure}

\section{Change in Mixing Entropy with Local Radius}

\begin{figure}[H]
 \centering
 \includegraphics[width=1\textwidth]{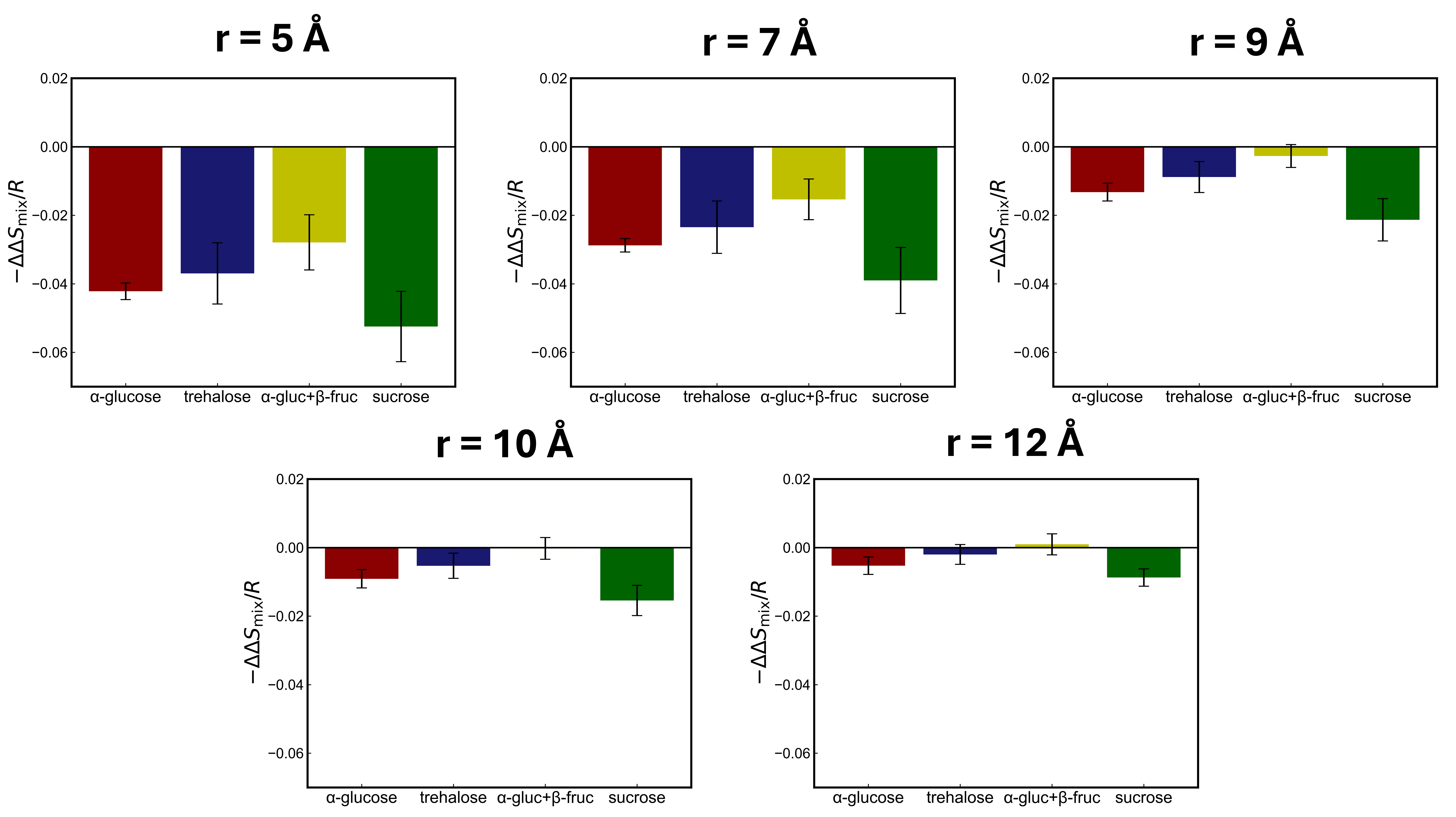}
 \caption{Change in the local mixing entropy with local radius around each bead of the HP in 2 M monosaccharide or 1 M disaccharide solutions. Positive values indicate higher mixing entropy in the folded state, suggesting enhanced configurational diversity of the local solvent environment.}
    \label{fig:si-delta-entropy}
\end{figure}

\section{Local Mixing Entropy with HP Radius of Gyration}

\begin{figure}[H]
 \centering
 \includegraphics[width=0.75\textwidth]{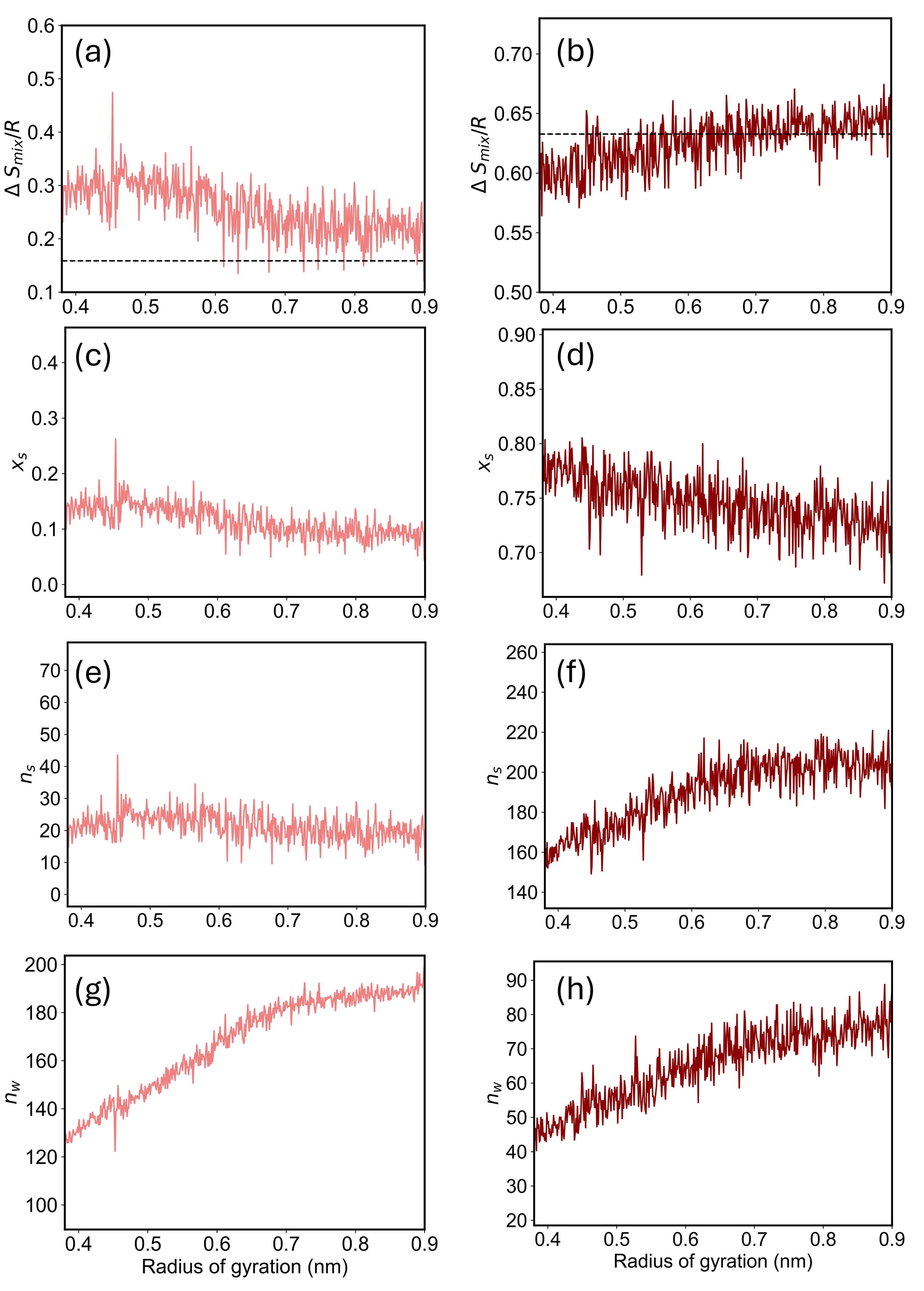}
 \caption{Change in (a,b) local mixing entropy ($\Delta S_{mix}$), (c,d) sugar atom number fraction ($x_s$), (e,f) number of local sugar atoms ($n_s$), and (g,h) number of local water atoms ($n_w$) with radius of gyration of the HP in 0.25 M (left, lighter shade) and 2 M (right, darker shade) $\alpha$-glucose solutions. Higher mixing entropy values indicate enhanced configurational diversity of the local solvent environment. The black dashed lines in the entropy plots indicate the mixing entropy of a perfectly mixed solution.}
    \label{fig:si-entropy-rg}
\end{figure}

\section{PMF Decomposition with Replicate Runs}

\begin{figure}[H]
 \centering
 \includegraphics[width=0.7\textwidth]{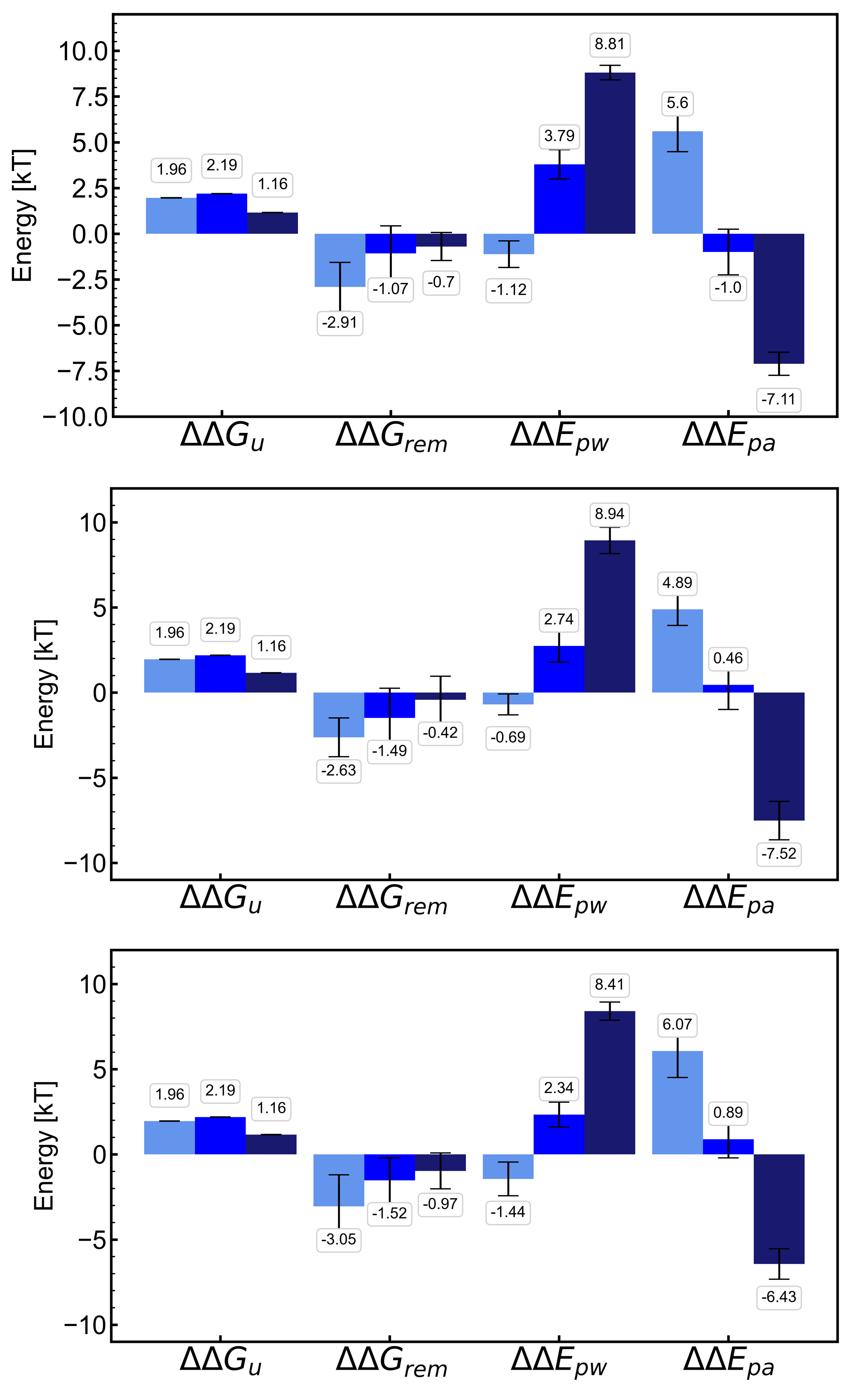}
 \caption{PMF Decomposition of 3 replicate trehalose runs at different concentrations (0.25 M, 0.5 M, 1 M). Increasing shading corresponds to increasing concentration of excipients.}
\label{fig:si-pmf-decomp-replicate}
\end{figure}

\section{Subgroup Interaction Energies of 2 M Solutions}

\begin{figure}[H]
 \centering
 \includegraphics[width=1\textwidth]{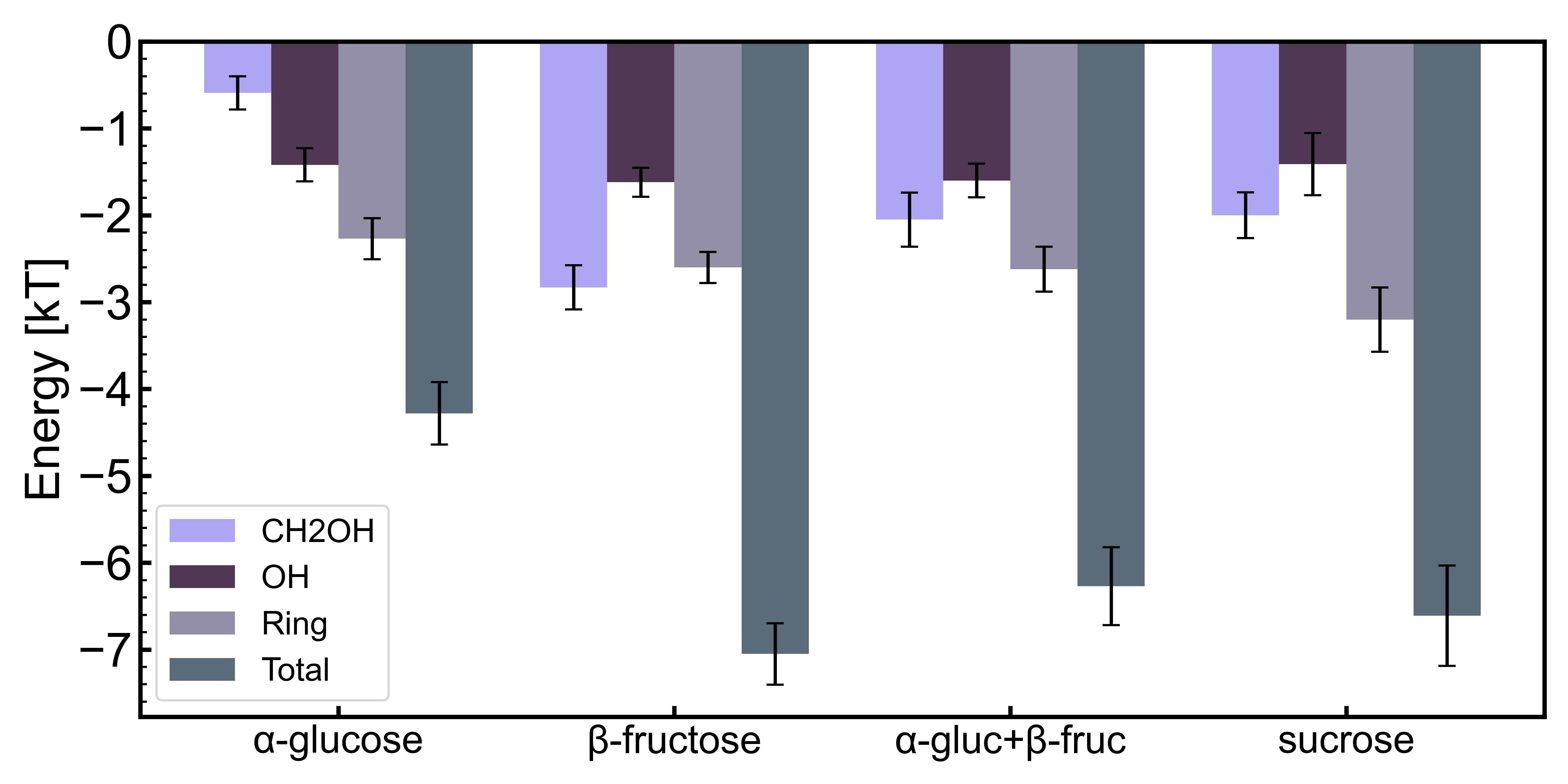}
 \caption{Difference in interaction energies of components with HP's unfolded and folded states for 2 M $\beta$-fructose, 1 M $\alpha$-glucose + 1 M $\beta$-fructose, and 1 M sucrose. }
    \label{fig:si-scrs-mono-finecomp}
\end{figure}

The $\alpha$-glucose-$\beta$-fructose mixture exhibited interaction signatures reminiscent of both $\alpha$-glucose and $\beta$-fructose: (1) ring groups had the dominant contribution to unfolding, similar to $\alpha$-glucose, and (2) the $\mathrm{CH_2OH}$ group also had a significant contribution, akin to $\beta$-fructose. The overall interaction energy magnitudes in the mixture aligned closely with those observed in 1 M sucrose and followed the same order: $\mathrm{OH} < \mathrm{CH_2OH} < \text{Ring}$ (Fig. \ref{fig:si-scrs-mono-finecomp}).

\section{CP Unfolding Thermodynamic Decomposition}

\begin{figure}[H]
 \centering
 \includegraphics[width=1\textwidth]{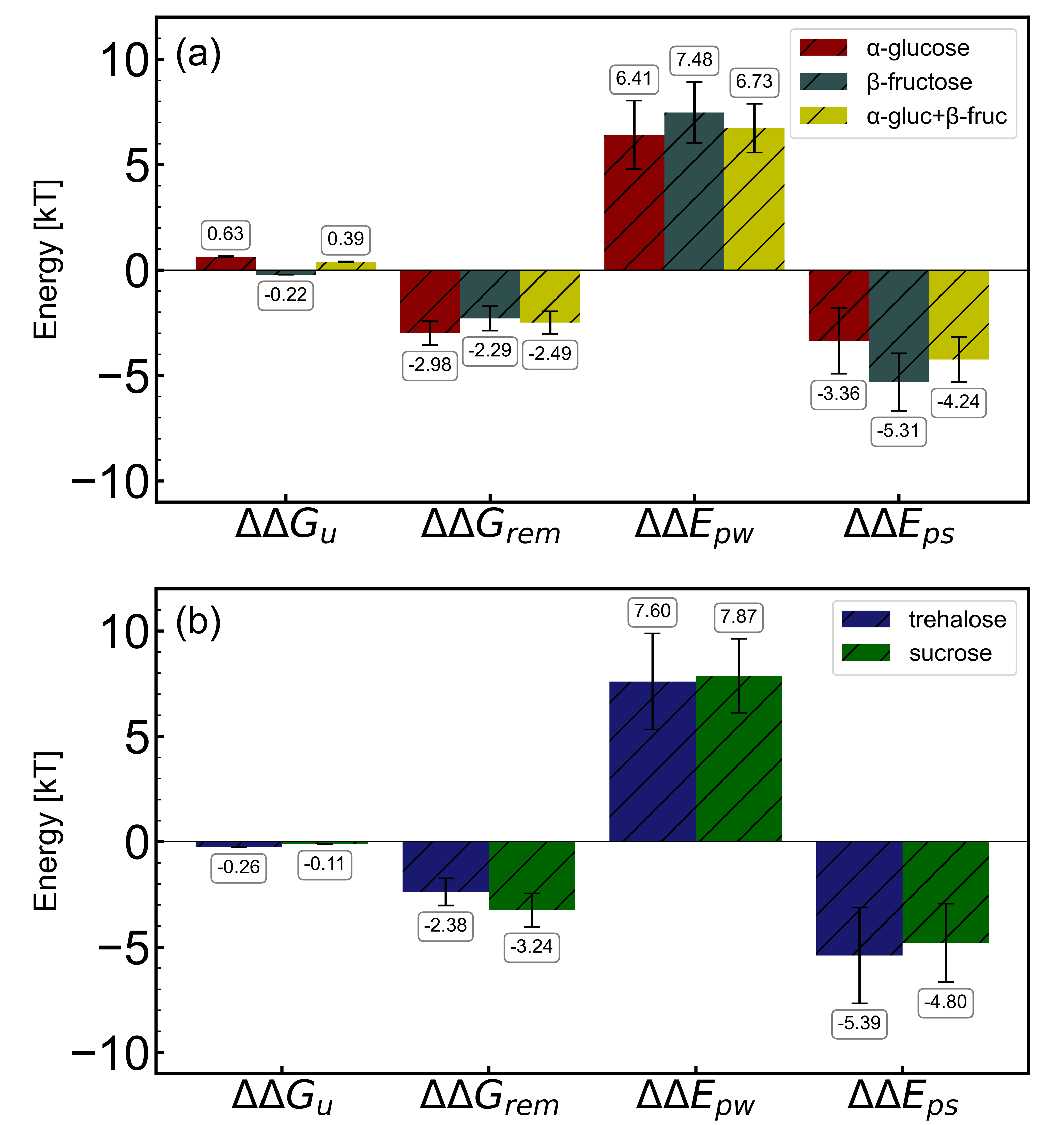}
 \caption{Decomposition of CP unfolding free energy contributions in (a) 2M monosaccharide ($\alpha$-glucose, $\beta$-fructose) or (b) 1M disaccharide (trehalose, sucrose) solutions. Positive
values favor folding and negative values favor unfolding of the CP.}
    \label{fig:cp-exc-pmf-decomp-mono-di}
\end{figure}

\clearpage
\bibliography{refs}